\documentclass[lettersize,journal]{IEEEtran}
\usepackage{amsmath,amsfonts}
\usepackage{algorithmic}

\usepackage{algorithm}
\usepackage{array}
\usepackage[caption=false,font=normalsize,labelfont=sf,textfont=sf]{subfig}
\usepackage{textcomp}
\usepackage{stfloats}
\usepackage{url}
\usepackage{verbatim}
\usepackage{graphicx}
\usepackage{cite}
\usepackage{adjustbox}
\usepackage{multirow}
\usepackage{amssymb,mathtools}
\usepackage{makecell} 
\usepackage{threeparttable}
\usepackage{xurl}
\usepackage{utfsym}
\usepackage[colorlinks=true,linkcolor=blue]{hyperref}

\usepackage{xcolor}
\usepackage{subcaption}
\usepackage{graphicx}
\usepackage{booktabs}

\usepackage{subcaption}
\usepackage{graphicx}
\usepackage{booktabs}
\usepackage{amsmath}
\usepackage{textcomp}
\usepackage{rotating}
\def\BibTeX{{\rm B\kern-.05em{\sc i\kern-.025em b}\kern-.08em
    T\kern-.1667em\lower.7ex\hbox{E}\kern-.125emX}}
\usepackage{adjustbox}
\usepackage{multirow}
\usepackage{amssymb,mathtools}
\usepackage{makecell} 
\usepackage{threeparttable}
\usepackage{hyperref}
\usepackage{xurl}
\usepackage{utfsym}

\usepackage{array}
\newcolumntype{P}[1]{>{\centering\arraybackslash}p{#1}}
\newcolumntype{M}[1]{>{\centering\arraybackslash}m{#1}}
\usepackage{tikz}
\usepackage{xcolor}

\mathchardef\mhyphen="2D

\newcommand{\Alg}[1]{\ensuremath{\mathsf{#1}}}

\newcommand{\Dec}{\Alg{Dec}}

\newcommand{\ClientAction}[1]{ 
	\node[right] at (\InitX, \Y) {#1};
}
\newcommand{\ServerAction}[1]{
	\node[left] at (\RespX, \Y) {#1};
}

\newcommand{\AdversaryAction}[1]{
	\node at ($1/2*(\InitX, \Y)+1/2*(\RespX, \Y)$) {#1};
}
\newcommand{\ClientToServer}[3][->]{
	\NextLine[0.5]
	\draw[#1] (\ArrowLeft,\Y) -- node[above] {#2} node[below] {#3} (\ArrowRight,\Y) ;
	\NextLine[0.5]
}
\newcommand{\ServerToClient}[3][->]{
	\NextLine[0.5]
	\draw[#1] (\ArrowRight,\Y) -- node[above] {#2} node[below] {#3} (\ArrowLeft,\Y) ;
	\NextLine[0.5]
}
\newcommand{\ClientToAdversary}[3][->]{
	\NextLine[0.5]
	\draw[#1] (\ArrowLeft,\Y) -- node[above] {#2} node[below] {#3} (\ArrowCenter,\Y) ;
	\NextLine[0.5]
}
\newcommand{\ServerToAdversary}[3][->]{
	\NextLine[0.5]
	\draw[#1] (\ArrowRight,\Y) -- node[above] {#2} node[below] {#3} (\ArrowCenter,\Y) ;
	\NextLine[0.5]
}

\newcommand{\AdversaryToClient}[3][->]{
	\NextLine[0.5]
	\draw[#1] (\ArrowCenter,\Y) -- node[above] {#2} node[below] {#3} (\ArrowLeft,\Y) ;
	\NextLine[0.5]
}
\newcommand{\AdversaryToServer}[3][->]{
	\NextLine[0.5]
	\draw[#1] (\ArrowCenter,\Y) -- node[above] {#2} node[below] {#3} (\ArrowRight,\Y) ;
	\NextLine[0.5]
}

\newcommand{\NextLine}[1][1.0]{
	\pgfmathparse{\Y+#1}
	\edef\Y{\pgfmathresult}
}
\newcommand{\linkgame}[2]{\hyperref[#1]{G#2}}

\newcounter{Bdversary}

\newcommand\orcidicon[1]{\href{https://orcid.org/#1}{\mbox{\scalerel*{
\begin{tikzpicture}[yscale=-1,transform shape]
\pic{orcidlogo};
\end{tikzpicture}
}{|}}}}

\newcommand*{\addFileDependency}[1]{
  \typeout{(#1)}

  \IfFileExists{#1}{}{\typeout{No file #1.}}
}

\newif\iffullversion
\newif\iffullversion
\newif\ifsubmissionversion
\fullversiontrue
\submissionversionfalse

\usepackage{adjustbox}
\usepackage{tikz}

\tikzset{
  orcidlogo/.pic={
    \fill[orcidlogocol] svg{M256,128c0,70.7-57.3,128-128,128C57.3,256,0,198.7,0,128C0,57.3,57.3,0,128,0C198.7,0,256,57.3,256,128z};
    \fill[white] svg{M86.3,186.2H70.9V79.1h15.4v48.4V186.2z}
                 svg{M108.9,79.1h41.6c39.6,0,57,28.3,57,53.6c0,27.5-21.5,53.6-56.8,53.6h-41.8V79.1z M124.3,172.4h24.5c34.9,0,42.9-26.5,42.9-39.7c0-21.5-13.7-39.7-43.7-39.7h-23.7V172.4z}
                 svg{M88.7,56.8c0,5.5-4.5,10.1-10.1,10.1c-5.6,0-10.1-4.6-10.1-10.1c0-5.6,4.5-10.1,10.1-10.1C84.2,46.7,88.7,51.3,88.7,56.8z};
  }
}

\usetikzlibrary{svg.path}
\usetikzlibrary{calc}

\begin{document}

\title{Privacy-preserving Robotic-based \\Multi-factor Authentication Scheme for \\Secure Automated Delivery System}

\author{Yang Yang,~\IEEEmembership{Student Member,~IEEE,}
Aryan Mohammadi Pasikhani,~\IEEEmembership{Member,~IEEE,}\\
Prosanta Gope,~\IEEEmembership{Senior Member,~IEEE,}
Biplab Sikdar,~\IEEEmembership{Senior Member,~IEEE,}
}

\markboth{IEEE Transactions on Information Forensics and Security}%
{Shell \MakeLowercase{\textit{et al.}}:IEEE Transactions on Information Forensics and Security}


\maketitle

\begin{abstract}
Package delivery is a critical aspect of various industries, but it often incurs high financial costs and inefficiencies when relying solely on human resources. The last-mile transport problem, in particular, contributes significantly to the expenditure of human resources in major companies. Robot-based delivery systems have emerged as a potential solution for last-mile delivery to address this challenge. However, robotic delivery systems still face security and privacy issues, like impersonation, replay, man-in-the-middle attacks (MITM), unlinkability, and identity theft.In this context, we propose a privacy-preserving multi-factor authentication scheme specifically designed for robot delivery systems. Additionally, AI-assisted robotic delivery systems are susceptible to machine learning-based attacks (e.g. FGSM, PGD, etc.). We introduce the \emph{first} transformer-based audio-visual fusion defender to tackle this issue, which effectively provides resilience against adversarial samples. Furthermore, we provide a rigorous formal analysis of the proposed protocol and also analyse the protocol security using a popular symbolic proof tool called ProVerif and Scyther. Finally, we present a real-world implementation of the proposed robotic system with the computation cost and energy consumption analysis. Code and pre-trained models are available at: \url{https://drive.google.com/drive/folders/18B2YbxtV0Pyj5RSFX-ZzCGtFOyorBHil?usp=sharing}
\end{abstract}

\section{Introduction}
Over recent years, the pervasive expansion of online shopping and emergent industries has brought exponential growth in the delivery service sector. These autonomous machines are designed to streamline the transportation of goods, reducing human error, improving efficiency, and ultimately saving businesses time and money. Powerhouse e-commerce platforms like Amazon, JD.com, and Alibaba have developed robust package logistics systems. In this context, the delivery industry has become an essential component of the online shopping ecosystem, underpinning the growth of many economies worldwide. Delivery robot applications span various industries, including food services, retail, healthcare, and personal use. For instance, they can be utilised for door-to-door parcel delivery, food delivery from restaurants, transporting medication and supplies in hospitals, or carrying groceries for elderly individuals or those with mobility issues. By operating in structured environments such as warehouses or navigating through unstructured terrains in public areas, these robots enhance operational efficiency and reduce carbon emissions by substituting traditional fuel-consuming delivery methods. As our societies continue to grapple with challenges like labour shortage, increasing consumer demand, and the urgent need for sustainable solutions, the importance of delivery robots is more pronounced than ever.
As highlighted by financial reports, the substantial costs associated with delivery services present a significant concern. For example, SF Express, a prominent courier company in China, revealed that human transportation costs amounted to 46.04\% of its total overheads in 2021 \cite{sf}. Similarly, another Chinese delivery company, Yuantong Express, reported that 53.84\% of its total costs were attributed to human transportation during the first half of 2021 \cite{yuantong}. Globally, the story is similar; Amazon Inc. spent an astonishing 76.7 billion dollars on transportation in 2021 \cite{amazon}.
These numbers underline the central issue of last-mile transport, which has emerged as a crucial component in reducing transport costs. A study cited in \cite{last-mile-delivery} indicated that last-mile transport costs account for 53\% of total transport costs. Many of these expenses arise from the human labour and fuel consumption associated with delivery vehicles. 
The global shift toward e-commerce and the substantial growth of delivery services, coupled with the demand for safer delivery options due to the pandemic, has accelerated the need for efficient and secure robotic delivery systems. However, despite the promising potential, these systems must overcome significant challenges to ensure successful integration into the existing logistics infrastructure. The global pandemic COVID-19 has also catalysed an escalating demand for contactless deliveries. The World Health Organisation (WHO) underscores the importance of open spaces, safe distancing, and limited time spent with others as primary measures for protecting oneself from infectious diseases.
In response to these converging trends, unmanned delivery robots present a promising solution. In this regard, robotic delivery can substantially reduce human labour and energy costs and offer a more adaptable delivery model. Furthermore, by facilitating quick deliveries in open-air environments with zero human interaction, these robots can significantly reduce the risk of infections. However, despite these advantages, implementing robotic delivery systems is challenging. For instance, secure and efficient data exchange between the robot, server, and client is paramount to the system's operational integrity, thereby highlighting the importance of robust communication protocols. Additionally, ensuring tasks' safe and efficient completion hinges on successfully verifying and identifying the robots.

{\subsection{Desirable properties for robotic-based multi-factor authentication (MFA) system }}

Considering the aforementioned discussion, it is anticipated that any robotic delivery system should fulfil the following Desirable Properties (DPs):

\subsubsection{DP1}\textbf{Secure Authentication.} It is crucial to verify the users before handling the item in any delivery system, including a Robotic Delivery system.

\subsubsection{DP2}\textbf{Multi-Factor Security.} Implementing multi-factor authentication in a delivery system is crucial for enhancing security and minimizing the potential for unauthorized access or tampering. Combining multiple factors significantly reduces the risk of unauthorized individuals compromising the delivery system or tampering with packages. This multi-factor authentication process instils confidence and ensures that only authorized personnel can handle and interact with the items being delivered, thereby guaranteeing the safety and security of the entire delivery process.

\subsubsection{DP3}\textbf{Enhanced User Privacy.} In the context of the AI-assisted Robotic Delivery system, privacy considerations play a pivotal role. It is imperative for the system to safeguard user privacy, necessitating the encryption of messages transmitted from the client. Additionally, deep learning-based methods employed for identity verification must effectively address privacy concerns, particularly about using users' facial and vocal biometric information. Biometric data, such as facial images and voiceprints, possess inherent attributes of being highly personal and distinctive identifiers, enabling accurate identification. However, the sensitive nature of these biometric data raises significant apprehensions regarding the privacy and security of individuals.

\subsubsection{DP4}\textbf{Advanced Face and Voice Verification model.} The deep learning model should be reliable and efficient for an AI-assisted Robotic Delivery system. This can be achieved by using the advanced deep learning model. Advanced deep learning models (such as Arcface and ECAPA-TDNN) would provide higher accuracy and stronger generalization capabilities, improving overall system performance. Higher accuracy helps prevent errors or unsuccessful verifications during the authentication process. Better generalization performance ensures adaptability to various scenarios and environments.

\subsubsection{DP5}\textbf{Resilience Against ML-based Adversary.} AI-assisted Robotic Delivery systems should have the ability to deal with ML-based attacks. The significance of ML-based attacks on AI-assisted Robotic Delivery systems lies in their potential to exploit vulnerabilities within the system's deep learning algorithms and compromise its operations. These attacks can have detrimental consequences,  posing risks to the robotic delivery system's reliability, security, and overall functionality.

\subsection{Related Work, Motivation, and Contribution of this article}

Robotic-based delivery is getting more attention in both the industry and academia. A considerable amount of research has been carried out in this area. For instance,  Amazon deployed its Scout system, an unmanned delivery robot \cite{scout} in March 2019. JD.COM also completed and deployed the first delivery of a robot back in 2017 \cite{jd_robot}. Terminus \cite{terminus} proposed their Titan AI robot for delivery and service. StarShip \cite{starship} is a company for food delivery robots and service robots. TuSimple \cite{tusimple} builds a framework for autonomous trucking. On the other hand, some notable academic research has also been carried out in this area. For instance, Bakach et al. \cite{Bakach} proposed a two-tier urban delivery network with robot-based deliveries; it can save 70\% of operational costs. Ostermeier \cite{Ostermeier} proposed cost-optimal routing for last-mile deliveries. There are also robots in many other fields, such as the medical industry. medRobo \cite{9441692} is a multi-functional medical robot that transports medicines and checks patients' physical indicators. Manikandan et al. \cite{9404698} have proposed intelligent nurse robots that can independently monitor and transport medicines consumption. Yang et al. \cite{YANG2021107779} proposed a secure shipping infrastructure using the delivery robot. They use QR codes and Person re-identification for delivery and authentication. Wang et al. \cite{9732339} proposed an AI-driven system for robot delivery; face verification and voice verification methods are proposed for robot delivery. However, protocol and artificial intelligence security still have some challenges and problems. {Meanwhile, several multi-factor authentication schemes were proposed in the literature. Shirvanian et al. \cite{shirvanian2014two} proposed a two-factor authentication scheme which can resist server compromise. MPCAuth \cite{tan2023mpcauth}, proposed by Tan et al., introduces a multi-factor authentication for distributed-trust systems. Zhang et al. 
\cite{zhang2017efficient} proposed a multi-factor authentication key exchange solution for mobile communication. Li et al. \cite{li2021practical} introduce threshold multi-factor authentication using fingerprint. Elliptic curve cryptography-based multi-factor authentication \cite{shukla2023design} was also proposed by Shukla1 et al. in 2023. Nevertheless, while several studies in the literature have proposed multi-factor authentication methods, none of them were designed considering robotic-based delivery systems.}

{\emph{\textbf{Tradiaional MFA vs Robotic-based MFA.}} In a traditional multi-factor authentication system, human interaction is typically limited to the user inputting credentials (such as a username and password). These systems rely heavily on the user's direct computer or mobile device engagement. Conversely, a robotic-based authentication system enhances this interaction by incorporating robots equipped with advanced sensors and machine-learning algorithms. These robots can engage with users in more intuitive and dynamic ways. Another difference is that traditional MFA generally requires a connection to a central server or a cloud service to verify user credentials. This dependency on online verification can be a limitation in environments where internet connectivity is unreliable. On the other hand, the robotic-based MFA designed in this article can run in an offline environment; using traditional MFA could increase the potential risk of leaking authentication keys, hence all data stored in the robot. Using multi-modal biometric-based MFA can prevent the need to store the user's credentials and prevent the risk of leaking data.}

\begin{table*}[ht]
\caption{SUMMARY OF THE RELATED WORK}
\centering

\scalebox{0.9}{
\centering
\begin{tabular}{ c | l | l | c c c c c}

\toprule[1.5pt]
\multirow{2}{*}{\makebox[0.22\textheight]{\textbf{Multi-Factor Authentication (MFA) Settings}}} &\multirow{2}{*}{\makebox[0.15\textwidth]{\textbf{Scheme}}} &\multirow{2}{*}{\makebox[0.23\textwidth]{\textbf{Authentication Techniques Used}}} & \multicolumn{5}{c}{\makebox[0.13\textwidth]{\textbf{Supported Desirable Properties}}} \\

{} &{} &  {} &{\textbf{DP1}} &{\textbf{DP2}} & {\textbf{DP3}} & {\textbf{DP4}} & {\textbf{DP5}}  \\

\midrule[1.5pt]

\multirow{5}{*}{\makebox[0.15\textwidth]{\textbf{Traditional MFA}}} & Tan et al. \cite{tan2023mpcauth} & Crypto + user profile & \usym{1F5F8} & \usym{1F5F8} & \usym{2613} & \usym{2613} & \usym{2613} \\
\cline{2-8}
{} & Li et al. \cite{li2021practical} & Crypto and Fingerprint & \usym{1F5F8} & \usym{1F5F8} & \usym{2613} & \usym{2613} & \usym{2613} \\
\cline{2-8}
{} & Zhang et al. \cite{zhang2017efficient} & Crypto and Fingerprint & \usym{1F5F8} & \usym{1F5F8} & \usym{2613} & \usym{2613} & \usym{2613} \\
\cline{2-8}
{} & Shirvanian et al. \cite{shirvanian2014two} & Crypto-based & \usym{1F5F8} & \usym{1F5F8} & \usym{2613} & \usym{2613} & \usym{2613} \\
\cline{2-8}
{} & Shukla et al. \cite{shukla2023design} & Crypto and Fingerprint & \usym{1F5F8} & \usym{1F5F8} & \usym{2613} & \usym{2613} & \usym{2613} \\

\midrule[1.5pt]

\multirow{7}{*}{\makebox[0.15\textwidth]{\textbf{Robotic-based MFA}}} &Manikandan et al. \cite{9404698}& RFID-based &\usym{1F5F8} & \usym{2613}& \usym{2613} & \usym{2613} & \usym{2613}  \\
\cline{2-8}
{} &Joy et al. \cite{9441692}& RFID-based & \usym{1F5F8}& \usym{2613} & \usym{2613} & \usym{2613} & \usym{2613}  \\
\cline{2-8}
{} &Jain et al. \cite{jain2021open} & Third party authentication&\usym{1F5F8}& \usym{2613} & - & \usym{2613} & \usym{2613}  \\
\cline{2-8}
{} &Liang et al. \cite{liang2021secure}& Crypto-based &\usym{1F5F8}  & \usym{2613} & \usym{2613}  & \usym{2613} & \usym{2613}  \\
\cline{2-8}
{} &\multirow{1}{*}{Yang et al. \cite{YANG2021107779} } & Crypto-based \textbf{or} Single-Biometric  & \multirow{1}{*}{\usym{1F5F8}} & \multirow{1}{*}{ \usym{2613} }& \multirow{1}{*}{ \usym{2613}   }& \multirow{1}{*}{ \usym{2613}} & \multirow{1}{*}{ \usym{2613}} \\

\cline{2-8}

{} &Wang et al. \cite{9732339} &  Biometric &  \usym{1F5F8} & \usym{2613} & \usym{2613} & \usym{2613} & \usym{2613}\\
\cline{2-8}

{} &\multirow{1}{*}{\textbf{Proposed Scheme} } & Crypto-based \textbf{and} Multimodel Biometrics & \multirow{1}{*}{ \usym{1F5F8}} & \multirow{1}{*}{ \usym{1F5F8} }& \multirow{1}{*}{ \usym{1F5F8}  }& \multirow{1}{*}{ \usym{1F5F8}} & \multirow{1}{*}{ \usym{1F5F8}} \\

\bottomrule[1.5pt]

\end{tabular}
}

\begin{tablenotes}
        \footnotesize
        \item \textbf{DP1:} Secure Authentication; \textbf{DP2:} Multi-Factor Security; \textbf{DP3:} Enhanced User's Privacy; \textbf{DP4:} Advanced Face and Voice Verification Model; \textbf{DP5:} Resilience Against ML-based Adversary; \textbf{Multimodal Biometrics :} With multiple biometric identifiers (face and voice).
\end{tablenotes}
\label{DPs}
      \vspace{-6mm} 

\end{table*}

\emph{\textbf{Motivation:}} Existing literature in the field of robot delivery has made numerous attempts to develop authentication protocols, but none have successfully fulfilled the requirements of a secure, multi-factor, privacy-preserving scheme. Furthermore, the proposed robot delivery schemes have failed to address challenges like noisy environments and machine learning-based adversarial samples. Therefore, it is imperative for a robot delivery authentication scheme to tackle these security concerns in real-world scenarios. A robust authentication scheme is vital for safeguarding user property and navigating complex interactive environments in any delivery scenario. To bridge this critical gap in the field of robotic delivery, we present a secure privacy-preserving multi-factor framework that specifically addresses issues related to model robustness, communication security, and the presence of machine learning-based attacks in existing robotic-based delivery systems. The \textit{major contributions} of this article are summarized as follows.

\begin{itemize}
    \item We propose the \textbf{\emph{first robotic-based multi-factor authentication protocol for robotic delivery systems}}, which pioneers the integration of cryptographic security with deep learning-based verification. One notable property of the proposed scheme is that it can support user registration under the insecure channel.

    \item We provide a \textbf{comprehensive security analysis(Using Reduction Proof and Symbolic Tool)} of our proposed scheme, where we considered both the crypto-based adversary model and AI-enabled adversary. The crypto-based adversary model evaluates the \textbf{key indistinguishability security, unlinkability, and perfect forward secrecy (PFS)}, showing our proposed scheme can be secure against impersonation, replay attacks, MITM (man-in-the-middle attacks), etc. attacks. Additionally, we present experimental results that demonstrate the efficacy of our defence mechanism in mitigating the impact of adversarial samples against the AI-enabled adversary.

     \item In order to tackle \textbf{AI-generated adversarial samples}, we have proposed \textbf{a groundbreaking defence mechanism known as the Audio-Visual Fusion Denoise Transformer Defense.} This innovative solution is the first of its kind, designed specifically to address the challenges posed by adversarial samples in the audio-visual domain.

    \item Our proposed scheme is able to \textbf{protect the user's biometric information}. Our proposed scheme does not require storing the user's biometric information for authentication in plain text. Still, we store the user's encrypted embedding rather than voice and face data for biometric authentication using the key generated in our proposed protocol.

    \item Our proposed scheme is implemented and thoroughly tested on a \textbf{real-world robotic platform}, specifically a Turtlebot3 integrated with Raspberry Pi 3. The programming of the robot component is conducted using Python 3.5.2 and PyTorch 1.12.0, ensuring a reliable and efficient execution of the proposed scheme.

\end{itemize}

\subsection{Paper Organisation}

The rest of the article is organized as follows: Section \ref{sec:Preliminaries} briefly introduces the elliptic curve, AI-based user verification and Adversarial samples. In Section \ref{sec:sys-dg-am}, we present our system architecture, design goals and adversary model. Section \ref{sec:Proposed} presents our secure multi-factor authentication protocol. In section \ref{sec:fusion}, we discuss our proposed audio-visual fusion denoise transformer. Section \ref{sec:discussion} presents experiment results and discussion. Finally, section \ref{sec:Conclusion} gives a conclusion about the paper.

\section{Preliminaries}
\label{sec:Preliminaries}

This section provides a concise overview of the elliptic curve cryptography employed in our protocol. Furthermore, we present some preliminary concepts related to deep learning, which serve as the foundation for the deep learning-based techniques in our protocol. 

\subsection{Cryptographic Notions}
Here we first provide some cryptographic notions for our protocol.

\subsubsection{Elliptic Curve Cryptography}

\

\textbf{Elliptic Curve Cryptography (ECC)} is a branch of public-key cryptography that utilizes the mathematical properties of elliptic curves for secure cryptographic operations \cite{ecc1,ecc2}. An elliptic curve can be defined as:

\begin{equation}
\label{ec}
y^2 = x^3 + ax + b
\end{equation}where $a$ and $b$ are constants. Points on the curve can be combined using specific mathematical operations, such as point addition and doubling, forming a group structure.

The \textbf{Elliptic Curve Diffie-Hellman (ECDH)} protocol is a cryptographic mechanism that facilitates secure key exchange between two parties by utilizing elliptic curves \cite{ecdh}. By leveraging the mathematical properties of elliptic curves, the ECDH protocol enables the establishment of a shared secret key over an insecure communication channel. The ECDH protocol can be formally defined as a tuple of algorithms, which includes \textbf{Key Generation} and \textbf{Key Arrangement}:

\begin{itemize}
    \item \textbf{Key Generation} is a key generation algorithm for the Elliptic Curve Diffie-Hellman protocol; it generates the secret key and public key respectively: $sk\ \leftarrow_\$ \textbf{KGen}(i^\lambda)$, $pk\ \leftarrow sk*G$.

    \item \textbf{Key Arrangement} is the algorithm that facilitates the computation of the shared secret key: $key = sk_c * pk_s$
\end{itemize}

The \textbf{Elliptic Curve Digital Signature Algorithm (ECDSA)} is a cryptographic scheme that utilizes elliptic curves to provide secure digital signatures \cite{ecdsa}. It offers essential security features such as data integrity, authentication, and non-repudiation in secure communication protocols. The ECDSA protocol can be described as a collection of algorithms that work together to ensure the integrity and authenticity of digital signatures:

\begin{itemize}
    \item \textbf{Key Generation} is a key generation algorithm for the Elliptic Curve Digital Signature Algorithm; it generates the secret key and public key for digital signature, respectively: $sk\ \leftarrow_\$ \textbf{KGen}(i^\lambda)$, $pk\ \leftarrow sk*G$.

    \item \textbf{Sign Signature} algorithm utilizes the secret key to create a digital signature for a given message: $ \sigma: z = hash(m), S_1 = k^{-1} * (h + sk * R) \ mod  \ p, \sigma \leftarrow (R, S)$.

    \item \textbf{Verify Signature} algorithm allows the recipient to verify the authenticity and integrity of the received message by using the sender's public key and the generated signature: $z = hash(m), P=S^{-1}*z*G + S^{-1}*R*pk$, \textbf{Check?} $P.x = R$.
\end{itemize}

\subsection{Deep-learning based Face and Voice Verification}

For deep learning-based face verification and voice verification, the more common approach is to extract the user's face representation and voice representation through a neural network, also known as face embedding or voice embedding. Two different training methods exist a Softmax classification network based on supervised learning to extract the embedding and a triplet-loss-based method based on self-supervised learning. Both methods can perform well. AAMSoftmax (ArcFace) \cite{arc_face} proposed by Deng et al. in 2019, has now been widely applied to face embedding and voice embedding, which can provide more accurate classification than normal Softmax and thus extract more accurate feature representations. The loss function of AAM-Softmax can be explained as follows:

\begin{equation}
\label{AAM}
\mathcal{L}_{AAM-Softmax} = -log \frac{ e^{s\ cos(\theta_{y_i}+m)} }{ e^{s\ cos(\theta_{y_i}+m)} + \Sigma^N_{j = i,j \neq y_i} e^{s\ cos(\theta_{j})}}
\end{equation}

where $\theta$ is the angle between the weight and feature, and $s$ is the radius of the hypersphere from learned embedding features. $m$ is an additive angular margin penalty.

\subsection{Adversarial Samples and Adversarial Training}

Szegedy et al. \cite{szegedy2013intriguing} in 2014 proposed that many deep neural networks are vulnerable to adversarial sample attacks. Applying a small perturbation to the original input data would make the model's output incorrect. Since then, more and more attack algorithms have been proposed in different ways. Also, existing literature has proposed that the model's defence against adversarial samples can be enhanced by adversarial training. Here, we briefly introduce three typical algorithms we use in this article.

\subsubsection{The Fast Gradient Sign Method}
The Fast Gradient Sign Method (FGSM) was proposed by Goodfellow \cite{goodfellow2014explaining} in 2015 and it is a gradient-based attack. It obtains an optimal max-norm constrained perturbation of $P$, equation \ref{fgsm} shows the details. 

\begin{equation}
\label{fgsm}
P = \alpha sign (\nabla_x J(\Theta, x, y))
\end{equation}

where $x, y$ is the original input image and label, $\Theta$ is model parameter and $J$ is loss function.

\subsubsection{Projected Gradient Descent}
Projected Gradient Descent (PGD) was proposed by Madry et al. \cite{pgd} with a simple scheme for maximizing the inner part of the saddle point formulation. They use a multi-step variant of the negative loss function to get more powerful adversarial samples. Equation \ref{pgd} shows more details of the loss function.
\begin{equation}
\label{pgd}
x^{t+1}=\Pi_{x+\mathcal{S}}(x^t+\alpha sign(\nabla_x J(\Theta, x, y)))
\end{equation}

\subsubsection{Basic Iterative Method}
The basic Iterative Method (BIM) or Iterative Fast Gradient Sign Method (I-FGSM) was proposed by Kurakin et al. \cite{bim} as an extension of the Projected Gradient Descent (PGD). It allows the attacker to apply FGSM multiple times with a suitable step size. With more iterations, this attack algorithm can produce harder samples for the deep learning model. Equation \ref{bim} shows more details of the algorithm.

\begin{equation}
\label{bim}
X^{adv}_0 = X,\ X^{adv}_{N+1}=Clip_X\{ X^{adv}_N + \alpha sign(\nabla_x J(X^{adv}_N, y_{true})) \}
\end{equation}

where $X^{adv}_{N}$ is the $N$th adversarial sample and $Cilp$ is a function to make the image in RGB format.

\section{System Architecture, Design Goals and Adversary Model}
\label{sec:sys-dg-am}

In this section, we first introduce our multi-factor authentication delivery system architecture. Subsequently, we describe the design security goals based on the implementation scenarios. Finally, we describe the two different adversary models.

\subsection{System Architecture}

\begin{figure*}[t!]
\center{\includegraphics[scale=0.20,trim=20 10 10 100,clip]{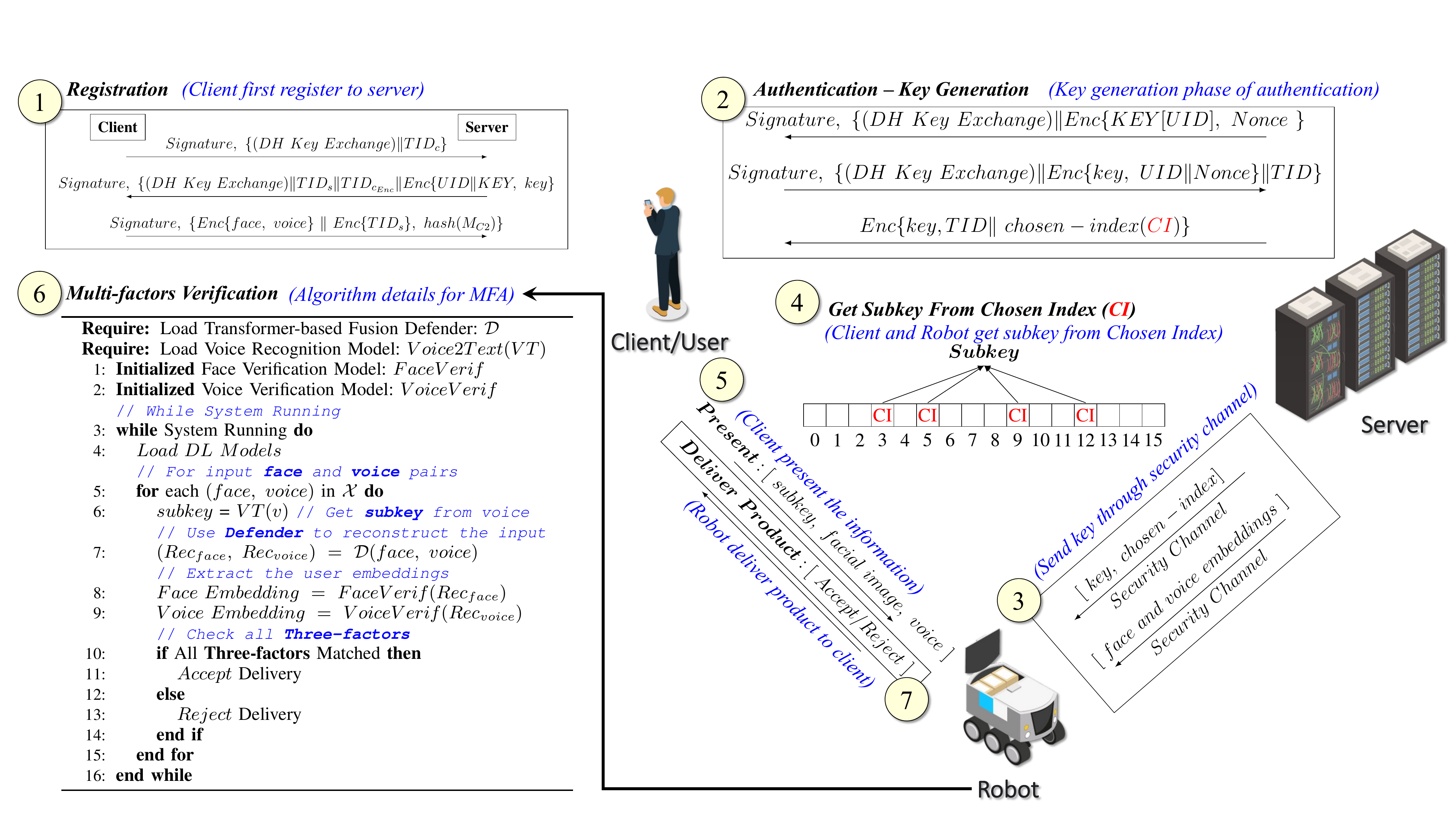}}
\caption{System Model and Overview of Our Proposed Robotic-based Multi-Factor Authentication Scheme.}
\label{fig:system_model} 
      \vspace{-6mm} 
\end{figure*}

Figure \ref{fig:system_model} shows our system model for the robotic delivery system. Our system model consists of three major entities: a server (located at the collection point), a client (user) and a robot. The server is responsible for the secret key exchange and the robot system initialisation in our proposed system. A client can order online and securely receive his/her package by providing his/her security credentials and other factors (such as facial image and voice) for authentication. In our proposed system, a robot plays a major role by securely delivering the package to a legitimate client. The server and client communicate through the Internet, and the robot loads the machine learning model from the secure channel.

The overall system is divided into two parts. The first part is the initialisation step that takes place after the arrival of the package—the system exchanges data with the client based on this part's package information. A random security code is generated with the client. At the same time, the machine learning model and the security code are loaded into the robot using the internal security network. The second part is when the robot arrives at the destination. Upon arrival at the destination, the transport robot enters the identification phase. During this phase, the delivery robot continuously captures images and sounds of its surroundings, waiting for potential users or capturing speech signals with a wake-up command. After receiving the appropriate information, the robot moves from the non-cooperative user identification phase to the cooperative user verification phase.

\subsection{Adversary Model}
In our adversary model, we consider both protocol-level (\textbf{Type 1}) and AI-enabled adversaries (\textbf{Type 2}). The abilities of each adversary are discussed as follows. 

\begin{itemize}
    \item In our adversary model, a \textbf{Type 1} adversary 
    is allowed to capture the messages communicated between the entities through the public channel. Then, the adversary may also try to alter and break the confidentiality of those messages. Our Type 1 adversary may also try to intercept, delete, insert and modify any message. Since users need to send their face and voice data through the public channel, the adversary may also try to break User privacy. Meanwhile, adversaries also try to break user anonymity by using linkability attacks.

    \item In our \textbf{Type 2} adversary, we consider an AI-enabled attacker who may try to break a deep learning-based verification model by generating a perturbation based on the original image or voice. We analyse five different attack algorithms on deep neural networks, which are FGSM \cite{goodfellow2014explaining}, PGD \cite{pgd}, BIM \cite{bim}, FFGSM \cite{ffgsm} and Jitter \cite{jitter}. These models are widely used and have a huge impact on the deep learning-based model; in order to defend against this type of adversary, we proposed our Audio-Visual Fusion Transformer.

\end{itemize}

\subsection{Design Goals}


 The newly developed Robot Delivery Protocol supports package delivery scenarios. Although it can meet all functional requirements, several requirements must be fulfilled to maintain security and privacy in the robot delivery system. In this regrade, our proposed protocol considered multiple design goals, which were discussed later.

 \subsubsection{\textbf{Mutual authentication with Multi-Factor Security}}Ensuring the authenticity of server, client, and robot is essential for robotic-based delivery. Hence, the server and robot communicate in a secure channel; mutual authentication is only required between the server and the client. In order to deliver the package, the client and robot also need to authenticate.

 \subsubsection{\textbf{Enhanced User's privacy}}Privacy preservation aims to protect sensitive data from malicious attackers. Our package delivery protocol uses a different approach containing both cryptography and biometric authentication. However, the user's face and voice information is private data that must be protected. In this regard, we have considered privacy-preserving properties as one of our design goals. 

 \subsubsection{\textbf{Strong anonymity}}Apart from the privacy of users' sensitive data, it is also important to anonymise user identities while transmitting the package. In order to avoid the adversary linking the steps of protocol execution, we need strong anonymity with unlinkability \cite{tifs_r}. Our scheme can maintain unlinkability and the details are provided in Section \ref{sec:sec_ana}.

  \subsubsection{\textbf{Perfect forward secrecy (PFS)}}PFS is essential in the protocol to ensure that if the current secret key has been compromised, there is no influence on the previous key. In the registration phase, users must send their face and voice information to the server. In order to maintain data privacy, all sensitive information is encrypted by shared keys. PFS ensure that if the current key leaks, there is no influence on the previous message containing face and voice data. Ephemeral Diffie-Hellman key exchange with authentication allows our scheme to address the PFS.

 \subsubsection{\textbf{Resilience against adversarial samples}}In a robot delivery system, where autonomous robots navigate and transport goods, ensuring the system's security and protection against adversarial samples is paramount. Adversarial samples are intentionally crafted inputs designed to deceive or manipulate deep learning models. To address the vulnerabilities associated with adversarial samples, it is essential to consider them as design goals. Specifically, the system should exhibit robustness, which entails its ability to accurately and reliably accomplish the tasks of item transportation and identity verification, even in the presence of adversarial inputs.

\section{Proposed Secure Robotic-based Multi-Factor Authentication Protocol}
\label{sec:Proposed}
This section first gives a high-level overview of our proposed multi-factor protocol with face and voice verification. Subsequently, we provide the details of each phase of the protocol.

\subsection{Hign-level Overview of Our Proposed Protocol}
\vspace{-1mm} 

Now, we present a comprehensive overview (also shown in Figure \ref{fig:system_model}) of our proposed protocol. As discussed in Section \ref{sec:sys-dg-am}, our system model comprises three major entities: the User, Server, and Robot. The primary objective of the proposed protocol is to establish a secure authentication mechanism between the User and the Robot during package delivery. To accomplish this objective, we have divided our protocol into two distinct phases: registration and authentication. Recognizing the insufficiency of single-factor authentication in ensuring user security, we have incorporated deep learning-based face and voice verification techniques within the protocol to enhance the overall security level. Consequently, we have further divided the authentication phase into two factors: the Crypto factor and the Biometric factor. To ensure better usability, our proposed secure authentication scheme mandates only a single interaction between the user and the robot, thereby enabling the robot to obtain all the necessary information. During the authentication process, the User needs to input the correct subkey (6 characters) generated by the Server from a secret key. Further details regarding this process will be provided subsequently. To commence our exposition, we first briefly introduce each phase of our proposed scheme.

The \textbf{Registration} phase is responsible for enrolling and incorporating new Clients and Robots into the system. During this phase, the Server generates initialization parameters for both the Client and the Robot. Concurrently, the Server collects the user's biometric information for verification purposes. To ensure the secure transmission of user data over the Internet, appropriate encryption techniques are employed. Upon receiving the user's information, including face images and voice recordings, the Server utilizes a deep learning model to extract corresponding face and voice embeddings. In order to address privacy concerns, the Server refrains from storing the actual face images and voice recordings, instead opting to retain only the extracted face and voice embeddings obtained through the use of the verification neural network.

The subsequent phase, termed the \textbf{Authentication} phase, encompasses generating secret codes and user authentication. A visual representation of the message flow within this phase is depicted in Figure \ref{fig:authentication}. This phase necessitates the involvement of all network components, namely the Client, Server, and Robot. Commencing with the initiation of the delivery process by the Server upon receiving the package ordered by the Client, a shared key is established between the Server and the Client, employing a key exchange protocol. Upon receiving feedback from the Client, the Server generates a secret key for authentication. Furthermore, the server randomly generates an index corresponding to the secret key and a sub
\subsection{Registration Phase}

In the registration phase of our proposed protocol, we assume that the communication channel between the Robot and Server is secure, given the Robot's standby at the collection point within the company. During the execution of this phase, the Server undertakes the generation of essential credentials for both the new Robot and new Client, which encompass the information of unique identifiers such as PID and TID. Simultaneously, the Client is responsible for collecting the user's face and voice information, which will be used for deep learning-based user authentication. To ensure a structured approach, our registration phase is divided into two distinct components: Robot registration and Client Registration. The subsequent sections provide comprehensive details regarding each component.

\subsubsection{\textbf{Robot Registration}}To register a new Robot to the network, each robot should share its essential information with the Server via the secure channel. Upon receiving the Robot's information, the Server generates the PID, TID, and a long secret key, which are then stored in the local list. This protocol consists of the following steps and is illustrated in Fig \ref{fig:Registration-robot}:

\begin{itemize}

\item[]\textbf{Step $\mathbf{A_1}$:} $\mathbf{M_{RR_1}}$ :$ \{ System \ Parameter \} $.

When a new Robot tries to register with the Server, it first sends all system parameters to the Server through the secure channel.

\item[]\textbf{Step $\mathbf{A_2}$:} $\mathbf{M_{RR_2}}$ :$ \{ k,\ PID,\ TID, \} $.

\end{itemize}

After the Server receives the information from the robot, it randomly generates the long secret key $k,\ PID,\ TID$ and sends them to the Robot through the secure channel.key can be selected from it. Upon completing the generation process, the Server securely transmits the secret key and index to the Robot via a secure channel.

\iffullversion
\begin{figure}[t!]
	\centering
	\begin{adjustbox}{max width=0.5\textwidth, max height=1\textheight}
		\fbox{
	\begin{tikzpicture}[yscale=-0.55,>=latex]
    \tikzstyle{every node}=[font=\large]
	\edef\InitX{0}
	\edef\ArrowLeft{1}
	\edef\ArrowCenter{6}
	\edef\ArrowRight{11}
	\edef\RespX{12}
	\edef\Y{0}
	
    \node [rectangle,draw,inner sep=6pt,right] at (\InitX,\Y) {\textbf{Robot} };
    \node [rectangle,draw,inner sep=6pt] at (11,\Y) {\textbf{Server}};




    \NextLine[2]

    \ClientAction{\textbf{Init. } System}
    \ServerAction{\textbf{Init. } System}

    \NextLine[1.5]
    \ClientToServer{\framebox[1.1\width]{$\boldsymbol{M_{RR_1}}:[System \ Parameter]$}}{$Security \ Channel$}

    \NextLine[1.5]
    \ServerAction{$ k_i \leftarrow_\$ \{0,1\}^{\lambda}, PID \leftarrow_\$ \{0,1\}^{\lambda}, TID \leftarrow_\$ \{0,1\}^{\lambda} $}

    \NextLine[2]
    \ServerToClient{\framebox[1.1\width]{$\boldsymbol{M_{RR_2}}:[k_i, \ PID, \ TID]$}}{$Security \ Channel$}

    \NextLine[1.5]
    \ClientAction{\textbf{Store: }$k_i, \ PID_i, \ TID_i$}
    \ServerAction{\textbf{Store: }$k_i, \ PID_i, \ TID_i$}

	\end{tikzpicture}
}
	\end{adjustbox}	\caption{ Robot Registration. }
	\label{fig:Registration-robot}
  \vspace{-6mm} 

\end{figure}
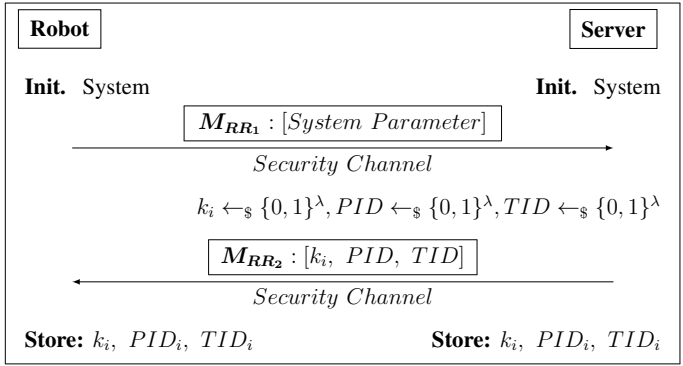
\else
\begin{figure}[htb]
	\centering
	\begin{adjustbox}{max width=1\textwidth, max height=1\textheight}
		\fbox{
	\begin{tikzpicture}[yscale=-0.55,>=latex]
    \tikzstyle{every node}=[font=\large]
	\edef\InitX{0}
	\edef\ArrowLeft{1}
	\edef\ArrowCenter{6}
	\edef\ArrowRight{11}
	\edef\RespX{12}
	\edef\Y{0}
	
    \node [rectangle,draw,inner sep=6pt,right] at (\InitX,\Y) {\textbf{Robot} };
    \node [rectangle,draw,inner sep=6pt] at (11,\Y) {\textbf{Server}};




    \NextLine[2]

    \ClientAction{\textbf{Init. } System}
    \ServerAction{\textbf{Init. } System}

    \NextLine[1.5]
    \ClientToServer{\framebox[1.1\width]{$\boldsymbol{M_{RR_1}}:[System \ Parameter]$}}{$Security \ Channel$}

    \NextLine[1.5]
    \ServerAction{$ k_i \leftarrow_\$ \{0,1\}^{\lambda}, PID \leftarrow_\$ \{0,1\}^{\lambda}, TID \leftarrow_\$ \{0,1\}^{\lambda} $}

    \NextLine[2]
    \ServerToClient{\framebox[1.1\width]{$\boldsymbol{M_{RR_2}}:[k_i, \ PID, \ TID]$}}{$Security \ Channel$}

    \NextLine[1.5]
    \ClientAction{\textbf{Store: }$k_i, \ PID_i, \ TID_i$}
    \ServerAction{\textbf{Store: }$k_i, \ PID_i, \ TID_i$}

	\end{tikzpicture}
}
	\end{adjustbox}
	\caption{ Registration: Robot. }
	\label{fig:Registration-robot}
 \vspace{-12mm} 
\end{figure}
\fi

\subsection{Registration Phase}

In the registration phase of our proposed protocol, we assume that the communication channel between the Robot and Server is secure, given the Robot's standby at the collection point within the company. During the execution of this phase, the Server undertakes the generation of essential credentials for both the new Robot and new Client, which encompass the information of unique identifiers such as PID and TID. Simultaneously, the Client is responsible for collecting the user's face and voice information, which will be used for deep learning-based user authentication. To ensure a structured approach, our registration phase is divided into two distinct components: Robot registration and Client Registration. The subsequent sections provide comprehensive details regarding each component.

\subsubsection{\textbf{Robot Registration}}To register a new Robot to the network, each robot should share its essential information with the Server via the secure channel. Upon receiving the Robot's information, the Server generates the PID, TID, and a long secret key, which are then stored in the local list. This protocol consists of the following steps and is illustrated in Fig \ref{fig:Registration-robot}:

\begin{itemize}

\item[]\textbf{Step $\mathbf{A_1}$:} $\mathbf{M_{RR_1}}$ :$ \{ System \ Parameter \} $.

When a new Robot tries to register with the Server, it first sends all system parameters to the Server through the secure channel.

\item[]\textbf{Step $\mathbf{A_2}$:} $\mathbf{M_{RR_2}}$ :$ \{ k,\ PID,\ TID, \} $.

After the Server receives the information from the robot, it randomly generates the long secret key $k,\ PID,\ TID$ and sends them to the Robot through the secure channel.

\end{itemize}

\subsubsection{\textbf{Client Registration}}

Before initiating the delivery order, each client is required to complete the registration process and exchange necessary registration information with the server. It is assumed that both the Client and Server possess knowledge of each other's public keys beforehand. Upon receipt of a client's registration request, the server undertakes the verification of the client's digital signature and subsequently employs a key exchange protocol to generate a secret key for secure information transmission. The user's face and voice information are subjected to processing by the server utilizing a deep learning model, resulting in the extraction of face embeddings and voice embeddings. The protocol entails a sequence of steps visually represented in Figure \ref{fig:Registration-client}.

\iffullversion
\begin{figure}[t!]
	\centering
	\begin{adjustbox}{max width=0.5\textwidth, max height=1\textheight}
		\fbox{
	\begin{tikzpicture}[yscale=-0.55,>=latex]
    \tikzstyle{every node}=[font=\large]
	\edef\InitX{0}
	\edef\ArrowLeft{1}
	\edef\ArrowCenter{6}
	\edef\ArrowRight{11}
	\edef\RespX{12}
	\edef\Y{0}
	
    \node [rectangle,draw,inner sep=6pt,right] at (\InitX,\Y) {\textbf{Client} };
    \node [rectangle,draw,inner sep=6pt] at (11,\Y) {\textbf{Server}};




    \NextLine[2]
    
    \ServerAction{\textbf{Init. } System}
    
    \NextLine
    \ServerAction{\textbf{ML:} Train Model}
    
    \NextLine
    \ClientAction{$d_c \leftarrow_\$ Z_q,\ G \in E_q(a, b),\ Q_c \leftarrow_\$ d_c * G,\ TID_c \leftarrow Z_q$}

    \NextLine
    \ClientAction{$M_{C1} = \{ Q_c \| TID_c \},\ z = hash(M_{C1})$}

    \NextLine
    \ClientAction{$k \leftarrow_\$ K,\ P = k * G,\ R_{C1} = P\ x\ label$}

    \NextLine
    \ClientAction{$S_{C1} = k^{-1} * (z + sk_c * R_{C1}) \ mod  \ p,\ \sigma_{c_1} \leftarrow (R_{C1}, S_{C1})$}


    
    \NextLine[1.5]
    \ClientToServer{\framebox[1.1\width]{$\boldsymbol{M_{RC_1}}:[\ \sigma_{c_1},\ M_{C1} ]$}}{}

    \NextLine[0.5]
    \ServerAction{ $z = hash(M_{C1}),\ P=S_{C1}^{-1}*z*G + S_{C1}^{-1}*R_{C1}*pk_c$  }
    
    \NextLine
    \ServerAction{\textbf{Check?} $P.x = R_{C1}$  }

    \NextLine
    \ServerAction{$d_s \leftarrow_\$ Z_q,\ G \in E_q(a, b),\ Q_s \leftarrow_\$ d_s * G,\ TID_s \leftarrow Z_q$}

    \NextLine
    \ServerAction{$key = KDF(d_s * Q_c),\ TID_{c_{Enc}} = Enc\{ TID_c,\ key \}$}

    \NextLine
    \ServerAction{$M_{S1} = \{ Q_s \| TID_s \| TID_{c_{Enc}} \|  Enc\{ UID\|KEY,\ key \} \}$}

    \NextLine
    \ServerAction{$z = hash(M_{S1}),\ k \leftarrow_\$ K,\ P = k * G,\ R_{S1} = P\ x\ label$}


    \NextLine
    \ServerAction{$S_{S1} = k^{-1} * (z + sk_s * R_{S1}) \ mod  \ p,\ \sigma \leftarrow (R_{S1}, S_{S1})$}

    \NextLine[1.5]
    \ServerToClient{\framebox[1.1\width]{$\boldsymbol{M_{RC_2}}:[\ \sigma,\ M_{S1} ]$}}{}

    \NextLine[0.5]
    \ClientAction{ $z = hash(M_{S1}), P=S_{S1}^{-1}*z*G + S_{S1}^{-1}*R_{S1}*pk_s$  }
    
    \NextLine
    \ClientAction{\textbf{Check?} $P.x = R_{S1}$  }

    \NextLine
    \ClientAction{$key = KDF(d_c * Q_S),\ TID_{{Dec}} = Dec\{ TID_{c_{Enc}},\ key \}$}

    \NextLine
    \ClientAction{\textbf{Check?} $TID_{Dec} = TID_c$  }


    \NextLine
    \ClientAction{$M_{C2} = \{ Enc\{ face,\ voice \}\ \| \ Enc\{ TID_s \},\ z = hash(M_{C2}) \}$}

    \NextLine
    \ClientAction{$k \leftarrow_\$ K,\ P = k * G,\ R_{C2} = P\ x\ label$}

    \NextLine
    \ClientAction{$S_{C2} = k^{-1} * (z + sk_c * R_{C2}) \ mod  \ p, \sigma\ \leftarrow (R_{C2}, S_{C2})$}

    \NextLine[1.5]
    \ClientToServer{\framebox[1.1\width]{$\boldsymbol{M_{RC_3}}:[\ \sigma,\ M_{C2} ] $}}{}

    \NextLine[0.5]
    \ServerAction{ $z = hash(M_{C2}),\ P=S_{C2}^{-1}*z*G + S_{C2}^{-1}*R_{C2}*pk_s$  }
    
    \NextLine
    \ServerAction{\textbf{Check?} $P.x = R_{C2}$  }

    \NextLine
    \ServerAction{ $TID_{Dec} = Dec\{ TID_{c_{Enc}},\ key\}$  }
    
    \NextLine
    \ServerAction{\textbf{Check?} $TID_{Dec} = TID$  }

    \NextLine
    \ServerAction{ $Face\ ,Voice\ \leftarrow \Dec\{ Enc\{ face,\ voice \} \}$  }

    \NextLine
    \ServerAction{\textbf{ML:} Get Embedding}


	
	\end{tikzpicture}
}
	\end{adjustbox}	\caption{ Client Registration }
	\label{fig:Registration-client}
   \vspace{-6mm} 
\end{figure}
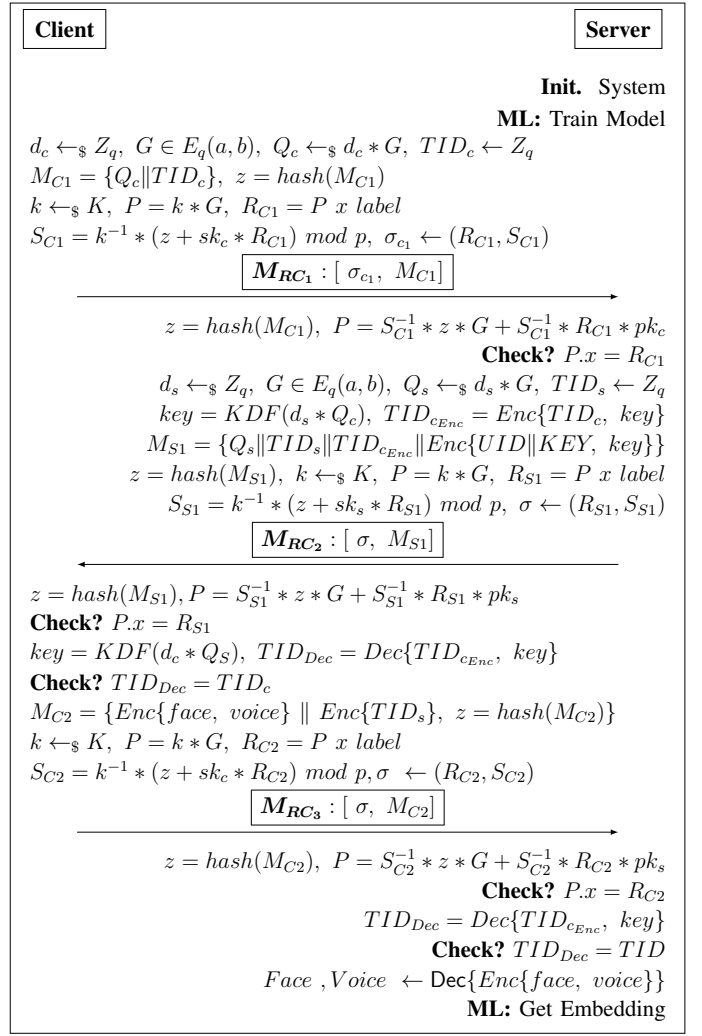
\else
\begin{figure}[htb]
	\centering
	\begin{adjustbox}{max width=1\textwidth, max height=1\textheight}
		\fbox{
	\begin{tikzpicture}[yscale=-0.55,>=latex]
    \tikzstyle{every node}=[font=\large]
	\edef\InitX{0}
	\edef\ArrowLeft{1}
	\edef\ArrowCenter{6}
	\edef\ArrowRight{11}
	\edef\RespX{12}
	\edef\Y{0}
	
    \node [rectangle,draw,inner sep=6pt,right] at (\InitX,\Y) {\textbf{Client} };
    \node [rectangle,draw,inner sep=6pt] at (11,\Y) {\textbf{Server}};




    \NextLine[2]
    
    \ServerAction{\textbf{Init. } System}
    
    \NextLine
    \ServerAction{\textbf{ML:} Train Model}
    
    \NextLine
    \ClientAction{$d_c \leftarrow_\$ Z_q,\ G \in E_q(a, b),\ Q_c \leftarrow_\$ d_c * G,\ TID_c \leftarrow Z_q$}

    \NextLine
    \ClientAction{$M_{C1} = \{ Q_c \| TID_c \},\ z = hash(M_{C1})$}

    \NextLine
    \ClientAction{$k \leftarrow_\$ K,\ P = k * G,\ R_{C1} = P\ x\ label$}

    \NextLine
    \ClientAction{$S_{C1} = k^{-1} * (z + sk_c * R_{C1}) \ mod  \ p,\ \sigma_{c_1} \leftarrow (R_{C1}, S_{C1})$}


    
    \NextLine[1.5]
    \ClientToServer{\framebox[1.1\width]{$\boldsymbol{M_{RC_1}}:[\ \sigma_{c_1},\ M_{C1} ]$}}{}

    \NextLine[0.5]
    \ServerAction{ $z = hash(M_{C1}),\ P=S_{C1}^{-1}*z*G + S_{C1}^{-1}*R_{C1}*pk_c$  }
    
    \NextLine
    \ServerAction{\textbf{Check?} $P.x = R_{C1}$  }

    \NextLine
    \ServerAction{$d_s \leftarrow_\$ Z_q,\ G \in E_q(a, b),\ Q_s \leftarrow_\$ d_s * G,\ TID_s \leftarrow Z_q$}

    \NextLine
    \ServerAction{$key = KDF(d_s * Q_c),\ TID_{c_{Enc}} = Enc\{ TID_c,\ key \}$}

    \NextLine
    \ServerAction{$M_{S1} = \{ Q_s \| TID_s \| TID_{c_{Enc}} \|  Enc\{ UID\|KEY,\ key \} \}$}

    \NextLine
    \ServerAction{$z = hash(M_{S1}),\ k \leftarrow_\$ K,\ P = k * G,\ R_{S1} = P\ x\ label$}


    \NextLine
    \ServerAction{$S_{S1} = k^{-1} * (z + sk_s * R_{S1}) \ mod  \ p,\ \sigma \leftarrow (R_{S1}, S_{S1})$}

    \NextLine[1.5]
    \ServerToClient{\framebox[1.1\width]{$\boldsymbol{M_{RC_2}}:[\ \sigma,\ M_{S1} ]$}}{}

    \NextLine[0.5]
    \ClientAction{ $z = hash(M_{S1}), P=S_{S1}^{-1}*z*G + S_{S1}^{-1}*R_{S1}*pk_s$  }
    
    \NextLine
    \ClientAction{\textbf{Check?} $P.x = R_{S1}$  }

    \NextLine
    \ClientAction{$key = KDF(d_c * Q_S),\ TID_{{Dec}} = Dec\{ TID_{c_{Enc}},\ key \}$}

    \NextLine
    \ClientAction{\textbf{Check?} $TID_{Dec} = TID_c$  }


    \NextLine
    \ClientAction{$M_{C2} = \{ Enc\{ face,\ voice \}\ \| \ Enc\{ TID_s \},\ z = hash(M_{C2}) \}$}

    \NextLine
    \ClientAction{$k \leftarrow_\$ K,\ P = k * G,\ R_{C2} = P\ x\ label$}

    \NextLine
    \ClientAction{$S_{C2} = k^{-1} * (z + sk_c * R_{C2}) \ mod  \ p, \sigma\ \leftarrow (R_{C2}, S_{C2})$}

    \NextLine[1.5]
    \ClientToServer{\framebox[1.1\width]{$\boldsymbol{M_{RC_3}}:[\ \sigma,\ M_{C2} ] $}}{}

    \NextLine[0.5]
    \ServerAction{ $z = hash(M_{C2}),\ P=S_{C2}^{-1}*z*G + S_{C2}^{-1}*R_{C2}*pk_s$  }
    
    \NextLine
    \ServerAction{\textbf{Check?} $P.x = R_{C2}$  }

    \NextLine
    \ServerAction{ $TID_{Dec} = Dec\{ TID_{c_{Enc}},\ key\}$  }
    
    \NextLine
    \ServerAction{\textbf{Check?} $TID_{Dec} = TID$  }

    \NextLine
    \ServerAction{ $Face\ ,Voice\ \leftarrow \Dec\{ Enc\{ face,\ voice \} \}$  }

    \NextLine
    \ServerAction{\textbf{ML:} Get Embedding}


	
	\end{tikzpicture}
}
	\end{adjustbox}
	\caption{ Registration: Client }
	\label{fig:Registration-client}
   \vspace{-6mm} 
\end{figure}
\fi

\begin{itemize}
\item[]\textbf{Step $\mathbf{B_1}$:} $\boldsymbol{M_{RC_1}}:\{\ \sigma,\ M_{C1} = \{ Q_c \| TID_c \}\ \}$.

When a new Client tries to register on the Server, it first generates a secret key $d_c$ for key exchange. After that, the Client computes $Q_c = D_c * G$ for the public key of ECDH. Then, the client gets the signature by computing $R_{C1}$ and $S_{C1} = k^{-1} * (z + sk_c * R_{C1}) \ mod  \ p$. Finally, the client sends message $M_{RC_1}$ to the server.

\item[]\textbf{Step $\mathbf{B_2}$:} $\boldsymbol{M_{RC_2}}:\{\ \sigma,\ M_{S1} \}$.

After receiving the message, the Server first computes the hash value of the message $M_{C1}$ and then computes $P=S_{C1}^{-1}*z*G + S_{C1}^{-1}*R_{C1}*pk_c$. If the Server receives a valid digital signature, it generates a secret key $d_s$ for key exchange. After that, the server gets the shared key by computing $key = d_c * Q_S$. Meanwhile, the server generates a unique UID and KEY for the client. Then, the server encrypts the $TID$ and the $UID\|KEY$ with a shared key. After that, the server computes its own signature $R$ equal to the value x of the point $P = (k*G)$. Finally, the Server computes $S_2 = k^{-1} * (h + sk^2_s * R) \ mod  \ p$ and send a message $\boldsymbol{M_{RC_2}}$ to the Client.

\item[]\textbf{Step $\mathbf{B_3}$:} $\boldsymbol{M_{RC_3}}:\{\ \sigma,\ M_{C2} \}$.

Upon receiving the message from the server, the client first computes the hash value of the message $M_{RC_2}$.  After that, the Client computes $P=S_{S1}^{-1}*z*G + S_{S1}^{-1}*R_{S1}*pk_s$ and check if this is a valid digital signature. After verifying the signature, the shared secret key is calculated by $key = d_c * Q_s$. Meanwhile, the client decrypts the message to get $UID$ and $KEY$. After checking the TID, the client gets the face and voice information of the user and encrypts them by using the shared key. Then, the client computes its own signature $R$ which is equal to the value x of the point $P = (k*G)$. Finally, the Server computes $S_{C2} = k^{-1} * (h + sk_c * R_{C_2}) \ mod  \ p$ and sends a message $\boldsymbol{M_{RC_3}}$ to the Server.

\item[]\textbf{Step $\mathbf{B_4}$:} Server Extracts User Embedding : 

After the Server receives the encrypted message from the client, it can decrypt the message using the shared key. After that, the Server can use a deep learning model to extract the face embedding and voice embedding for verification. Once the Server gets the user embedding, the face and voice information will be deleted because of security concerns.

\end{itemize}

\subsection{Authentication Phase}

As discussed, the authentication phase $(\textbf{Step C})$ has been further subdivided into two parts: \textit{verification of the crypto factor ($\textbf{C}^{Crypto}$)} and \textit{verification of the biometric factor ($\textbf{C}^{Biometric}$)}. The primary objective of the crypto factor is to generate a secret key along with a random index for cryptographic authentication. To enhance the security strength, we aim to generate a lengthy key for authentication.  However, it is impractical for the user to input or vocalize a lengthy sequence of characters manually. Thus, we employ a secret key with a random index to facilitate user verification. In this approach, the user must select a specific index from the key, simplifying the authentication process. Concurrently, the authentication process incorporates face and voice verification techniques to bolster the overall security level. The protocol encompasses a sequence of steps, visually illustrated in Figure \ref{fig:authentication}.

\iffullversion
\begin{figure}[t!]
	\centering
	\begin{adjustbox}{max width=0.50\textwidth, max height=1\textheight}
		\fbox{
	\begin{tikzpicture}[yscale=-0.55,>=latex]
    \tikzstyle{every node}=[font=\large]
	\edef\InitX{0}
	\edef\ArrowLeft{1}
	\edef\ArrowCenter{6}
	\edef\ArrowRight{11}
	\edef\RespX{12}
	\edef\Y{0}
	
    \node [rectangle,draw,inner sep=6pt,right] at (\InitX,\Y) {\textbf{Client} };
	\node [rectangle,draw,inner sep=6pt,left] at ((6.5,\Y)  {\textbf{Robot}};
    \node [rectangle,draw,inner sep=6pt] at (11,\Y) {\textbf{Server}};

    \node at (0,39) [draw, minimum height=0.5cm, minimum width=0.5cm, inner sep=0pt, text width=0.5cm, align=center] {\textcolor{red}{\fontsize{12}{12}\selectfont  }};
    \node at (0.5,39) [draw, minimum height=0.5cm, minimum width=0.5cm, inner sep=0pt, text width=0.5cm, align=center] {\textcolor{blue}{\fontsize{12}{12}\selectfont }};
    \node at (1,39) [draw, minimum height=0.5cm, minimum width=0.5cm, inner sep=0pt, text width=0.5cm, align=center] {\textcolor{green}{\fontsize{12}{12}\selectfont  }};
    \node at (1.5,39) [draw, minimum height=0.5cm, minimum width=0.5cm, inner sep=0pt, text width=0.5cm, align=center] {\textcolor{red}{\fontsize{12}{12}\selectfont  CI}};
    \node at (2,39) [draw, minimum height=0.5cm, minimum width=0.5cm, inner sep=0pt, text width=0.5cm, align=center] {\textcolor{orange}{\fontsize{12}{12}\selectfont   }};
    \node at (2.5,39) [draw, minimum height=0.5cm, minimum width=0.5cm, inner sep=0pt, text width=0.5cm, align=center] {\textcolor{red}{\fontsize{12}{12}\selectfont CI}};
    \node at (3,39) [draw, minimum height=0.5cm, minimum width=0.5cm, inner sep=0pt, text width=0.5cm, align=center] {\textcolor{brown}{\fontsize{12}{12}\selectfont  }};
    \node at (3.5,39) [draw, minimum height=0.5cm, minimum width=0.5cm, inner sep=0pt, text width=0.5cm, align=center] {\textcolor{teal}{\fontsize{12}{12}\selectfont  }};
    \node at (4,39) [draw, minimum height=0.5cm, minimum width=0.5cm, inner sep=0pt, text width=0.5cm, align=center] {\textcolor{magenta}{\fontsize{12}{12}\selectfont  }};
    \node at (4.5,39) [draw, minimum height=0.5cm, minimum width=0.5cm, inner sep=0pt, text width=0.5cm, align=center] {\textcolor{red}{\fontsize{12}{12}\selectfont CI}};
    \node at (5,39) [draw, minimum height=0.5cm, minimum width=0.5cm, inner sep=0pt, text width=0.5cm, align=center] {\textcolor{gray}{\fontsize{12}{12}\selectfont  }};
    \node at (5.5,39) [draw, minimum height=0.5cm, minimum width=0.5cm, inner sep=0pt, text width=0.5cm, align=center] {\textcolor{red}{\fontsize{12}{12}\selectfont  }};
    \node at (6,39) [draw, minimum height=0.5cm, minimum width=0.5cm, inner sep=0pt, text width=0.5cm, align=center] {\textcolor{red}{\fontsize{12}{12}\selectfont  CI}};
    \node at (6.5,39) [draw, minimum height=0.5cm, minimum width=0.5cm, inner sep=0pt, text width=0.5cm, align=center] {\textcolor{blue}{\fontsize{12}{12}\selectfont  }};
    \node at (7,39) [draw, minimum height=0.5cm, minimum width=0.5cm, inner sep=0pt, text width=0.5cm, align=center] {\textcolor{green}{\fontsize{12}{12}\selectfont  }};
    \node at (7.5,39) [draw, minimum height=0.5cm, minimum width=0.5cm, inner sep=0pt, text width=0.5cm, align=center] {\textcolor{black}{\fontsize{12}{12}\selectfont  }};
   
    \node[align=center] at (3.75, 36.5) {$\boldsymbol{Subkey}$};

    \node[align=center] at (0, 40) {0};
    \node[align=center] at (0.5, 40) {1};
    \node[align=center] at (1, 40) {2};
    \node[align=center] at (1.5, 40) {3};
    \node[align=center] at (2, 40) {4};
    \node[align=center] at (2.5, 40) {5};
    \node[align=center] at (3, 40) {6};
    \node[align=center] at (3.5, 40) {7};
    \node[align=center] at (4, 40) {8};
    \node[align=center] at (4.5, 40) {9};
    \node[align=center] at (5, 40) {10};
    \node[align=center] at (5.5, 40) {11};
    \node[align=center] at (6, 40) {12};
    \node[align=center] at (6.5, 40) {13};
    \node[align=center] at (7, 40) {14};
    \node[align=center] at (7.5, 40) {15};


    \draw[->] (1.5,38.5) -- (3.75, 37);
    \draw[->] (2.5,38.5) -- (3.75, 37);
    \draw[->] (4.5,38.5) -- (3.75, 37);
    \draw[->] (6,38.5) -- (3.75, 37);




    \NextLine[2]

    \ServerAction{\textbf{Init. } System}
    \ClientAction{\textbf{Init. } Order}

    \NextLine




    \ServerAction{\textbf{Package Arrived}}

    \NextLine
    \ServerAction{\textbf{Get:} $UID$ from package information  }

    \NextLine
    \ServerAction{$d_s \leftarrow_\$ Z_q,\ G \in E_q(a, b),\ Q_s \leftarrow_\$ d_s * G,\ \lambda \leftarrow \{0,1\}^n$}

    \NextLine
    \ServerAction{$M_{S1} = \{ Q_s \| Enc\{KEY[UID],\ \lambda\ \} \},\ z = hash(M_{S1})$}

    \NextLine
    \ServerAction{$k \leftarrow_\$ K,\ P = k * G,\ R_{S1} = P\ x\ label$}

    \NextLine
    \ServerAction{$S_{S1} = k^{-1} * (z + sk_s * R_{S1}) \ mod  \ p,\ \sigma_{s_1} \leftarrow (R_{S1}, S_{S1})$}

    \NextLine[1.5]
    \ServerToClient{\framebox[1.1\width]{$\boldsymbol{M_{AU_1}}:[\ \sigma,\ M_{S1} ]$}}{}

    \NextLine[0.5]
    \ClientAction{ $z = hash(M_{S1}), P=S_{S1}^{-1}*z*G + S_{S1}^{-1}*R_{S1}*pk_s$  }

    \NextLine
    \ClientAction{\textbf{Check?} $P.x = R_{S1}$  }

    \NextLine
    \ClientAction{   $key = KDF(d_c * Q_s),\ \lambda\ \leftarrow Dec\{KEY[UID],\ M_{S1}\}$  }

    \NextLine
    \ClientAction{$d_c \leftarrow_\$ Z_q,\ G \in E_q(a, b),\ Q_c \leftarrow_\$ d_c * G,\ \mu \leftarrow \{0,1\}^n,\ TID \leftarrow Z_q$}

    \NextLine
    \ClientAction{$M_{C1} = \{ Q_c \| Enc\{key,\ UID\| \mu  \}\| TID \},\ z = hash(M_{C1})$}

    \NextLine
    \ClientAction{$k \leftarrow_\$ K,\ P = k * G,\ R_{C1} = P\ x\ label$}

    \NextLine
    \ClientAction{$S_{C1} = k^{-1} * (z + sk_c * R_{C1}) \ mod  \ p,\ \sigma_{c_1} \leftarrow (R_{C1}, S_{C1})$}
=
    \NextLine
    \ClientAction{\textbf{Update: }$ KEY[UID]^* = KEY[UID] \oplus \lambda,\ UID^* = UID \oplus \mu $}

    \NextLine[1.5]
    \ClientToServer{\framebox[1.1\width]{$\boldsymbol{M_{AU_2}}:[\ \sigma_{c_1},\ M_{C1} ]$}}{}

    \NextLine[0.5]
    \ServerAction{ $z = hash(M_{C1}),\ P=S_{C1}^{-1}*z*G + S_{C1}^{-1}*R_{C1}*pk_c$  }

    \NextLine
    \ServerAction{\textbf{Check?} $P.x = R_{C1}$  }

    \NextLine
    \ServerAction{   $key = d_s * Q_c$  }

    \NextLine
    \ServerAction{   $UID_{Dec}\|\mu \leftarrow Dec\{key,\ M_{C1}\}$  }

    \NextLine
    \ServerAction{\textbf{Check?} $UID_{Dec} = UID$  }

    \NextLine
    \ServerAction{ $index \leftarrow Z_q,\ M_{S2} \leftarrow Enc\{ key, TID\|index \}$  }

    \NextLine
    \ServerAction{\textbf{Update: }$ KEY[UID]^* = KEY[UID] \oplus \lambda,\ UID^* = UID \oplus \mu $}

    \NextLine[1.5]
    \ServerToClient{\framebox[1.1\width]{$\boldsymbol{M_{AU_3}}:[\ M_{S2} ]$}}{}

    \NextLine[0.5]
    \ClientAction{ $ TID_{Dec}\|index \leftarrow Dec\{ key,\ M_{S2}\} $  }

    \NextLine
    \ClientAction{\textbf{Check?} $TID_{Dec} = TID$  }

    \NextLine[1.5]
    \ServerToAdversary{\framebox[1.1\width]{$\boldsymbol{M_{AU_4}}:[\ key,\ chosen-index(CI)\ ]$}}{$Security\ Channel$}

    \NextLine[1.5]
    \ServerToAdversary{\framebox[1.1\width]{$\boldsymbol{Load}:[\ Verification \ Models\ ]$}}{$Security\ Channel$}

    \NextLine[4.5]





    \NextLine[1.5]
    \ClientToAdversary{\framebox[1.1\width]{$\boldsymbol{Present}:[\ subkey, facial\ image, voice \ ]$}}{$Camera\ and\ Microphone$}

    \NextLine[1]
    \AdversaryAction{\textbf{Check?} $subkey,\ chosen-index$}
    \NextLine
    \AdversaryAction{$(Rec_{face},\ Rec_{voice}) \ = \ Defender(face,\ voice)$}
    \NextLine
    
    \AdversaryAction{\textbf{Check?} $FaceVerif(Rec_{face})$$\ \ \ $ }
    \NextLine
    \AdversaryAction{\textbf{Check?} $VoiceVerif(Rec_{voice})$$\ \ $ }

    \NextLine[1.5]
    \AdversaryToClient{\framebox[1.1\width]{$\boldsymbol{Deliver\ Product}:[\ Accept/Reject \ ]$}}{}


    \NextLine[1]
    \AdversaryToServer{\framebox[1.1\width]{$\boldsymbol{Response}:[\ Accept/Reject \ ]$}}{$Security\ Channel$}




    




	\end{tikzpicture}
}
	\end{adjustbox}	\caption{ Authentication Phase of the Proposed Scheme.}
	\label{fig:authentication}
   \vspace{-6mm} 
\end{figure}
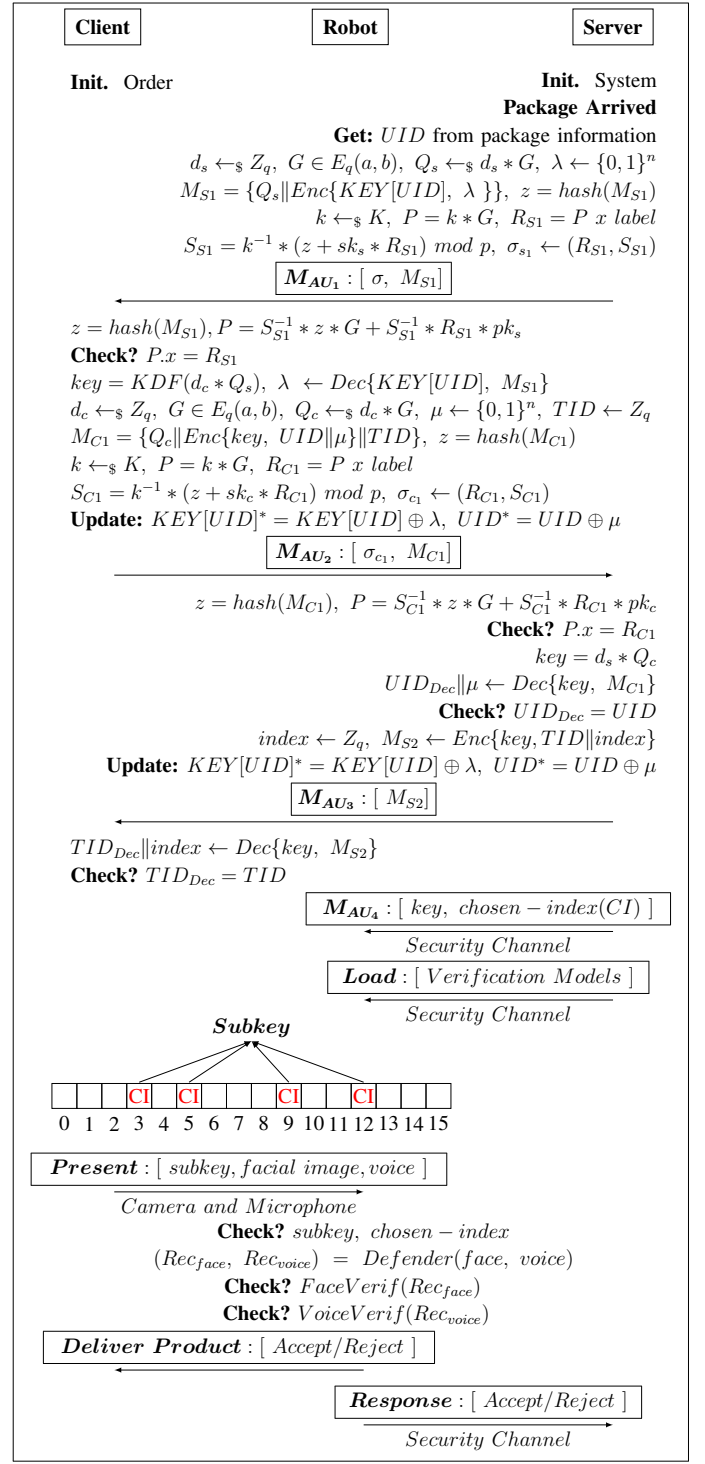
\else
\begin{figure}[htb]
	\centering
	\begin{adjustbox}{max width=1\textwidth, max height=1\textheight}
		\fbox{
	\begin{tikzpicture}[yscale=-0.55,>=latex]
    \tikzstyle{every node}=[font=\large]
	\edef\InitX{0}
	\edef\ArrowLeft{1}
	\edef\ArrowCenter{6}
	\edef\ArrowRight{11}
	\edef\RespX{12}
	\edef\Y{0}
	
    \node [rectangle,draw,inner sep=6pt,right] at (\InitX,\Y) {\textbf{Client} };
	\node [rectangle,draw,inner sep=6pt,left] at ((6.5,\Y)  {\textbf{Robot}};
    \node [rectangle,draw,inner sep=6pt] at (11,\Y) {\textbf{Server}};

    \node at (0,39) [draw, minimum height=0.5cm, minimum width=0.5cm, inner sep=0pt, text width=0.5cm, align=center] {\textcolor{red}{\fontsize{12}{12}\selectfont  }};
    \node at (0.5,39) [draw, minimum height=0.5cm, minimum width=0.5cm, inner sep=0pt, text width=0.5cm, align=center] {\textcolor{blue}{\fontsize{12}{12}\selectfont }};
    \node at (1,39) [draw, minimum height=0.5cm, minimum width=0.5cm, inner sep=0pt, text width=0.5cm, align=center] {\textcolor{green}{\fontsize{12}{12}\selectfont  }};
    \node at (1.5,39) [draw, minimum height=0.5cm, minimum width=0.5cm, inner sep=0pt, text width=0.5cm, align=center] {\textcolor{red}{\fontsize{12}{12}\selectfont  CI}};
    \node at (2,39) [draw, minimum height=0.5cm, minimum width=0.5cm, inner sep=0pt, text width=0.5cm, align=center] {\textcolor{orange}{\fontsize{12}{12}\selectfont   }};
    \node at (2.5,39) [draw, minimum height=0.5cm, minimum width=0.5cm, inner sep=0pt, text width=0.5cm, align=center] {\textcolor{red}{\fontsize{12}{12}\selectfont CI}};
    \node at (3,39) [draw, minimum height=0.5cm, minimum width=0.5cm, inner sep=0pt, text width=0.5cm, align=center] {\textcolor{brown}{\fontsize{12}{12}\selectfont  }};
    \node at (3.5,39) [draw, minimum height=0.5cm, minimum width=0.5cm, inner sep=0pt, text width=0.5cm, align=center] {\textcolor{teal}{\fontsize{12}{12}\selectfont  }};
    \node at (4,39) [draw, minimum height=0.5cm, minimum width=0.5cm, inner sep=0pt, text width=0.5cm, align=center] {\textcolor{magenta}{\fontsize{12}{12}\selectfont  }};
    \node at (4.5,39) [draw, minimum height=0.5cm, minimum width=0.5cm, inner sep=0pt, text width=0.5cm, align=center] {\textcolor{red}{\fontsize{12}{12}\selectfont CI}};
    \node at (5,39) [draw, minimum height=0.5cm, minimum width=0.5cm, inner sep=0pt, text width=0.5cm, align=center] {\textcolor{gray}{\fontsize{12}{12}\selectfont  }};
    \node at (5.5,39) [draw, minimum height=0.5cm, minimum width=0.5cm, inner sep=0pt, text width=0.5cm, align=center] {\textcolor{red}{\fontsize{12}{12}\selectfont  }};
    \node at (6,39) [draw, minimum height=0.5cm, minimum width=0.5cm, inner sep=0pt, text width=0.5cm, align=center] {\textcolor{red}{\fontsize{12}{12}\selectfont  CI}};
    \node at (6.5,39) [draw, minimum height=0.5cm, minimum width=0.5cm, inner sep=0pt, text width=0.5cm, align=center] {\textcolor{blue}{\fontsize{12}{12}\selectfont  }};
    \node at (7,39) [draw, minimum height=0.5cm, minimum width=0.5cm, inner sep=0pt, text width=0.5cm, align=center] {\textcolor{green}{\fontsize{12}{12}\selectfont  }};
    \node at (7.5,39) [draw, minimum height=0.5cm, minimum width=0.5cm, inner sep=0pt, text width=0.5cm, align=center] {\textcolor{black}{\fontsize{12}{12}\selectfont  }};
   
    \node[align=center] at (3.75, 36.5) {$\boldsymbol{Subkey}$};

    \node[align=center] at (0, 40) {0};
    \node[align=center] at (0.5, 40) {1};
    \node[align=center] at (1, 40) {2};
    \node[align=center] at (1.5, 40) {3};
    \node[align=center] at (2, 40) {4};
    \node[align=center] at (2.5, 40) {5};
    \node[align=center] at (3, 40) {6};
    \node[align=center] at (3.5, 40) {7};
    \node[align=center] at (4, 40) {8};
    \node[align=center] at (4.5, 40) {9};
    \node[align=center] at (5, 40) {10};
    \node[align=center] at (5.5, 40) {11};
    \node[align=center] at (6, 40) {12};
    \node[align=center] at (6.5, 40) {13};
    \node[align=center] at (7, 40) {14};
    \node[align=center] at (7.5, 40) {15};


    \draw[->] (1.5,38.5) -- (3.75, 37);
    \draw[->] (2.5,38.5) -- (3.75, 37);
    \draw[->] (4.5,38.5) -- (3.75, 37);
    \draw[->] (6,38.5) -- (3.75, 37);




    \NextLine[2]

    \ServerAction{\textbf{Init. } System}
    \ClientAction{\textbf{Init. } Order}

    \NextLine




    \ServerAction{\textbf{Package Arrived}}

    \NextLine
    \ServerAction{\textbf{Get:} $UID$ from package information  }

    \NextLine
    \ServerAction{$d_s \leftarrow_\$ Z_q,\ G \in E_q(a, b),\ Q_s \leftarrow_\$ d_s * G,\ \lambda \leftarrow \{0,1\}^n$}

    \NextLine
    \ServerAction{$M_{S1} = \{ Q_s \| Enc\{KEY[UID],\ \lambda\ \} \},\ z = hash(M_{S1})$}

    \NextLine
    \ServerAction{$k \leftarrow_\$ K,\ P = k * G,\ R_{S1} = P\ x\ label$}

    \NextLine
    \ServerAction{$S_{S1} = k^{-1} * (z + sk_s * R_{S1}) \ mod  \ p,\ \sigma_{s_1} \leftarrow (R_{S1}, S_{S1})$}

    \NextLine[1.5]
    \ServerToClient{\framebox[1.1\width]{$\boldsymbol{M_{AU_1}}:[\ \sigma,\ M_{S1} ]$}}{}

    \NextLine[0.5]
    \ClientAction{ $z = hash(M_{S1}), P=S_{S1}^{-1}*z*G + S_{S1}^{-1}*R_{S1}*pk_s$  }

    \NextLine
    \ClientAction{\textbf{Check?} $P.x = R_{S1}$  }

    \NextLine
    \ClientAction{   $key = KDF(d_c * Q_s),\ \lambda\ \leftarrow Dec\{KEY[UID],\ M_{S1}\}$  }

    \NextLine
    \ClientAction{$d_c \leftarrow_\$ Z_q,\ G \in E_q(a, b),\ Q_c \leftarrow_\$ d_c * G,\ \mu \leftarrow \{0,1\}^n,\ TID \leftarrow Z_q$}

    \NextLine
    \ClientAction{$M_{C1} = \{ Q_c \| Enc\{key,\ UID\| \mu  \}\| TID \},\ z = hash(M_{C1})$}

    \NextLine
    \ClientAction{$k \leftarrow_\$ K,\ P = k * G,\ R_{C1} = P\ x\ label$}

    \NextLine
    \ClientAction{$S_{C1} = k^{-1} * (z + sk_c * R_{C1}) \ mod  \ p,\ \sigma_{c_1} \leftarrow (R_{C1}, S_{C1})$}
=
    \NextLine
    \ClientAction{\textbf{Update: }$ KEY[UID]^* = KEY[UID] \oplus \lambda,\ UID^* = UID \oplus \mu $}

    \NextLine[1.5]
    \ClientToServer{\framebox[1.1\width]{$\boldsymbol{M_{AU_2}}:[\ \sigma_{c_1},\ M_{C1} ]$}}{}

    \NextLine[0.5]
    \ServerAction{ $z = hash(M_{C1}),\ P=S_{C1}^{-1}*z*G + S_{C1}^{-1}*R_{C1}*pk_c$  }

    \NextLine
    \ServerAction{\textbf{Check?} $P.x = R_{C1}$  }

    \NextLine
    \ServerAction{   $key = d_s * Q_c$  }

    \NextLine
    \ServerAction{   $UID_{Dec}\|\mu \leftarrow Dec\{key,\ M_{C1}\}$  }

    \NextLine
    \ServerAction{\textbf{Check?} $UID_{Dec} = UID$  }

    \NextLine
    \ServerAction{ $index \leftarrow Z_q,\ M_{S2} \leftarrow Enc\{ key, TID\|index \}$  }

    \NextLine
    \ServerAction{\textbf{Update: }$ KEY[UID]^* = KEY[UID] \oplus \lambda,\ UID^* = UID \oplus \mu $}

    \NextLine[1.5]
    \ServerToClient{\framebox[1.1\width]{$\boldsymbol{M_{AU_3}}:[\ M_{S2} ]$}}{}

    \NextLine[0.5]
    \ClientAction{ $ TID_{Dec}\|index \leftarrow Dec\{ key,\ M_{S2}\} $  }

    \NextLine
    \ClientAction{\textbf{Check?} $TID_{Dec} = TID$  }

    \NextLine[1.5]
    \ServerToAdversary{\framebox[1.1\width]{$\boldsymbol{M_{AU_4}}:[\ key,\ chosen-index(CI)\ ]$}}{$Security\ Channel$}

    \NextLine[1.5]
    \ServerToAdversary{\framebox[1.1\width]{$\boldsymbol{Load}:[\ Verification \ Models\ ]$}}{$Security\ Channel$}

    \NextLine[4.5]





    \NextLine[1.5]
    \ClientToAdversary{\framebox[1.1\width]{$\boldsymbol{Present}:[\ subkey, facial\ image, voice \ ]$}}{$Camera\ and\ Microphone$}

    \NextLine[1]
    \AdversaryAction{\textbf{Check?} $subkey,\ chosen-index$}
    \NextLine
    \AdversaryAction{$(Rec_{face},\ Rec_{voice}) \ = \ Defender(face,\ voice)$}
    \NextLine
    
    \AdversaryAction{\textbf{Check?} $FaceVerif(Rec_{face})$$\ \ \ $ }
    \NextLine
    \AdversaryAction{\textbf{Check?} $VoiceVerif(Rec_{voice})$$\ \ $ }

    \NextLine[1.5]
    \AdversaryToClient{\framebox[1.1\width]{$\boldsymbol{Deliver\ Product}:[\ Accept/Reject \ ]$}}{}


    \NextLine[1]
    \AdversaryToServer{\framebox[1.1\width]{$\boldsymbol{Response}:[\ Accept/Reject \ ]$}}{$Security\ Channel$}




    




	\end{tikzpicture}
}
	\end{adjustbox}
	\caption{Authentication }
	\label{fig:authentication}
   \vspace{-6mm} 
\end{figure}
\fi

\subsubsection{\textbf{Verification of the Crypto-factor of the Authentication Phase}}

The secret $Subkey$ and index generation process will be started after the server receives the user's package. After that, the server generates a long key for the $Subkey$ using a key exchange protocol. Meanwhile, the server randomly picks some of the digital in the key, named $Subkey$. Then, the server creates a digital signature for authentication. Finally, the server sends the information to both the robot and the client. Our proposed protocol consists of the following steps.

\begin{itemize}

\item[]\textbf{Step $\mathbf{C^{Crypto}_1}$:} $\boldsymbol{M_{AU_1}}:\{\ \sigma,\ M_{S1} \}$.

The server starts the robot delivery process when a new package is delivered at the collection point. It first gets the $UID$ from the package information. After that, the server calculates $Q_s = d_s * G$ for ECDH. Meanwhile, the server generates a value $\lambda$ for the KEY update. Then, the server computes the hash value of message $M_{S1} = \{ Q_s \| Enc\{KEY[UID],\ \lambda \} \}$. After that, the server computes the $R_{S1}$ equal to the value x of the point $P = (k*G)$. Finally, the server compute $S_{S1} = k^{-1} * (h + sk_s * R_{S1}) \ mod  \ p$ and sends a message $\boldsymbol{M_{S1}}$ to the client.

\item[]\textbf{Step $\mathbf{C^{Crypto}_2}$:} $\boldsymbol{M_{AU_2}}:\{\ \sigma,\ M_{C1} \}$.

After the client receives the message from the server, it first computes $ P=S_{S1}^{-1}*hash(M_{S1})*G + S_{S1}^{-1}*R_{S1}*pk_s $ and checks if this is a valid digital signature. After verifying the signature, the client computes the shared $key = d_c * Q_s$ and decrypts the message using the shared key to get the $\lambda$. After that, the client calculates $Q_c = d_c * G$ for ECDH. Meanwhile, the client generates a value $\mu$ for the KEY update and a TID for verification. Then, the client computes the hash value of message $ M_{C1} = \{ Q_c \| Enc\{KEY[UID],\ UID\| \mu \}\| TID \} $. After that, the client computes the $R_{C1}$ equal to the value x of the point $P = (k*G)$. Finally, the client computes $S_{C1} = k^{-1} * (h + sk_c * R_{C1}) \ mod  \ p$ and sends a message $\boldsymbol{M_{C1}}$ to the server.

\item[]\textbf{Step $\mathbf{C^{Crypto}_3}$:} $\boldsymbol{M_{AU_3}}:\{M_{S2} \}$.

Similarly, the server first computes $P=S_{C1}^{-1}*hash(M_{C1})*G + S_{C1}^{-1}*R_{C1}*pk_c$ to verify the digital signature. After that, the server generates the $index$ for the subkey and encrypts $TID\|index$ using the shared key. In the end, the server composes a $M_{S2}$ message, then sends the $M_{S2}$ to the client and updates the $UID$ and KEY.

\item[]\textbf{Step $\mathbf{C^{Crypto}_4}$:} Client Get $index$.

After receiving the message $M_{S2}$, the Client decrypts the message using the shared key and gets the index for the $subkey$.

\item[]\textbf{Step $\mathbf{C^{Crypto}_5}$:} $\boldsymbol{M_{AU_4}}:\{key,\ index,\ Embedding\}$.

In the meantime, the server sends the Client's biometric information (such as face and voice embedding) to the Robot through a secure channel. It contains the $key$ and $index$ information for crypto-based verification and the user embedding for deep learning-based verification.

\end{itemize}

\subsubsection{\textbf{Verification of the Biometric factor of the Authentication Phase }}

The verification process is the interaction between the User and the Robot. The User speaks out the correct code from the code matrix. At that time, the robot collects the face and voice information and extracts the user embedding. After that, the Robot checks the code, face embedding and voice embedding. If all information is checked, the delivery process will be finished. Our proposed protocol consists of the following steps.

\begin{itemize}

\item[]\textbf{Step $\mathbf{C^{Biometric}_1}$:} $\boldsymbol{MA_1}:\{Code,\ Face,\ Voice\}$.

After the robot reaches the client, users interact with the Robot and provide the essential information. While providing the secret subkey, the robot also records the voice and face information for verification. In order to protect against the adversarial samples, the input image and voice first feed into our proposed defender. The defender reconstructs the input data and then performs the verification. For more details of the defender, we provide in Section \ref{sec:fusion}.

\item[]\textbf{Step $\mathbf{C^{Biometric}_2}$:} $\boldsymbol{M_2}:\{Feedback\}$.

The Robot checks the user's identity according to the input information and finishes the delivery if all checks are successful. Note that the robot needs to check multiple factors for verification.

\item[]\textbf{Step $\mathbf{C^{Biometric}_3}$:} $\boldsymbol{M_3}:\{Return Back\}$.

The Robot finally goes back to the Robot station and sends back the delivery status.

\end{itemize}

As we discussed in Step $\mathbf{C^{Biometric}_1}$, we proposed a \emph{new} audio-visual fusion denoise transformer. Since there is no audio-visual-based fusion defender, we consider finding the relationship between the audio and face and using it for a new type of defender. An overview of our proposed model is depicted in Figure \ref{fig:Fusion}. In the next section, we give more details about our proposed model.

\section{Proposed Transformer-based Fusion Defender}
\label{sec:fusion}
In this section, we first give a high-level overview of our proposed Audio-Visual fusion denoise transformer. Subsequently, we provide the model structures and the training details.

\subsection{High level overview}

In order to achieve the desired security goals, particularly in the context of deep learning-based authentication models, we employ a defence model to mitigate the influence of AI-enabled attacks. Several approaches were proposed in the literature; autoencoder-based defender\cite{dae} used to be the most common way to protect the deep learning model by reconstructing the input data. Meanwhile, anomaly detection-based defenders can reject the data with high reconstruction errors. In our proposed model design, we follow the original Transformer \cite{transformer} and vision transformer \cite{vit} as closely as possible and combine face verification and voice verification together for enhanced security. Since face images are 2D data, we follow the way of face transformer \cite{face_transformer} to reshape the image into a sequence of flattened 2D patches. For audio data, we follow[paper] to extract the mel-spectrogram from the waveform record and reshape the mel-spectrogram into a sequence of flattened 2D patches.

Algorithm \ref{al:algorithm_ver} illustrates the details of how we implement our proposed fusion defender inside of the protocol.

\begin{figure*}[htb]
\center{\includegraphics[scale=0.40,trim=10 30 0 45,clip]{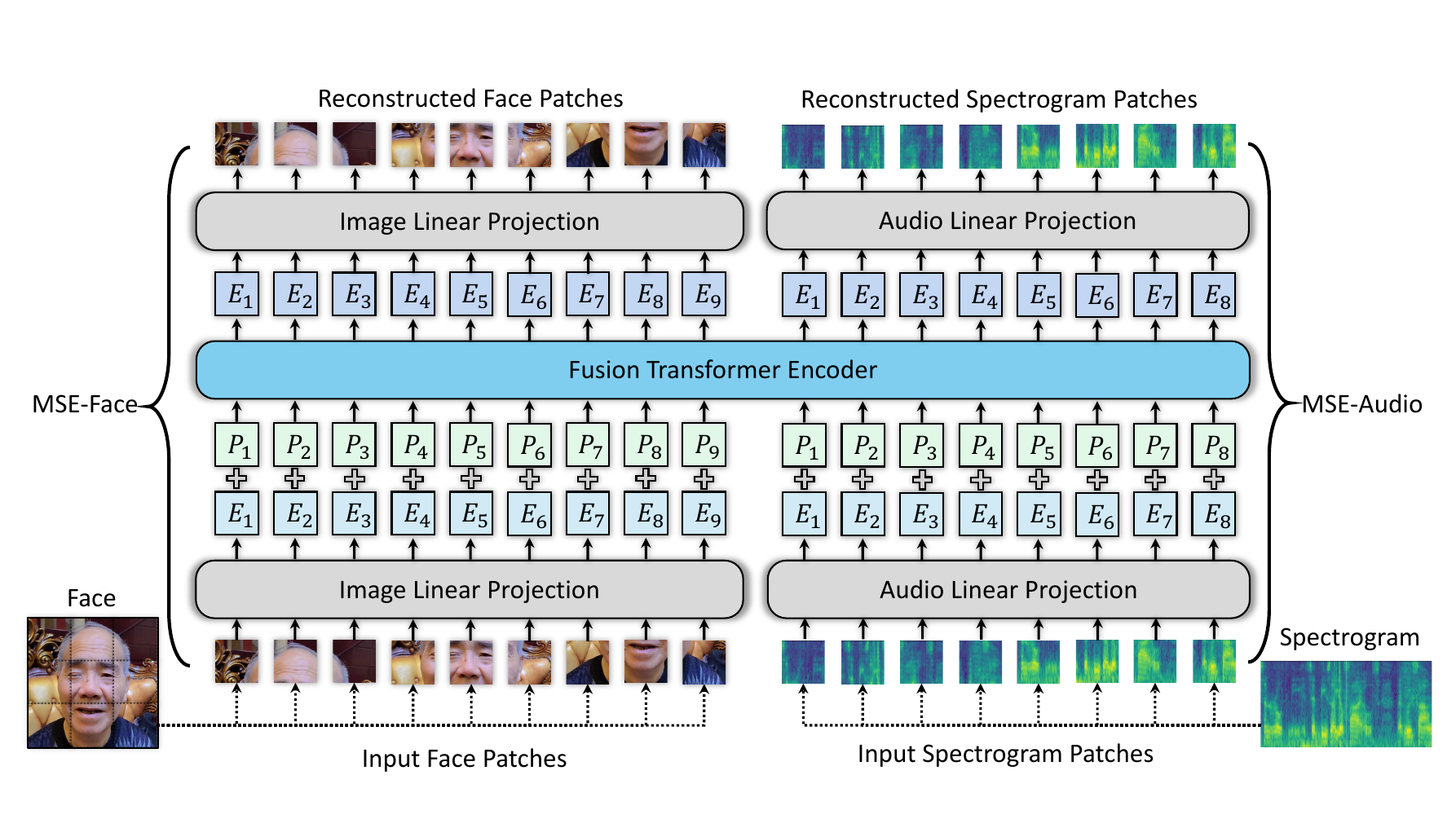}}
\caption{Proposed Audio-Visual Fusion Denoise Transformer.}
\label{fig:Fusion} 
      \vspace{-6mm} 
\end{figure*}

\begin{algorithm}[htp]
\caption{Verification Process with Defender}
\label{al:algorithm_ver}
\begin{algorithmic}[1]
\small

\REQUIRE \textbf{Load Transformer-based Fusion Defender:} $\mathcal{D}$\\
\REQUIRE \textbf{Load Voice Recognition Model:} $Voice2Text(VT)$\\

\STATE \textbf{Initialize Face Verification Model:} $FaceVerif$\\
\STATE \textbf{Initialize Voice Verification Model:} $VoiceVerif$\\
\STATE \textcolor{blue}{\textit{//While System Running}}
\WHILE{System Running} 
    
    \STATE \textcolor{blue}{\textit{//Load The DL Models}}
    \STATE Load DL Models

    \STATE \textcolor{blue}{\textit{//For input \textbf{face} and \textbf{voice} pairs}}
    \FOR{each $(face, voice)$ in $\mathcal{X}$}

        \STATE \textcolor{blue}{\textit{//Get subkey from voice}}
        \STATE $subkey = VT(voice)$

        \STATE \textcolor{blue}{\textit{//Use Defender to reconstruct the input}}
        \STATE $(Rec_{face}, Rec_{voice}) = \mathcal{D}(face, voice)$
        
        \STATE \textcolor{blue}{\textit{//Extract the user's Face embeddings}}
        \STATE $Face\ Embedding = FaceVerif(Rec_{face})$
        
        \STATE \textcolor{blue}{\textit{//Extract the user's Voice embeddings}}
        \STATE $Voice\ Embedding = VoiceVerif(Rec_{voice})$
        
        \STATE \textcolor{blue}{\textit{//Check All Three Factors}}
        \IF{All Three-factors Matched} 
            \STATE Accept Delivery
        \ELSE
            \STATE Reject Delivery
        \ENDIF

    \ENDFOR

\ENDWHILE
\end{algorithmic}
\end{algorithm}

\vspace{-6mm} 

\subsection{Methodology}

\subsubsection{Problem Definition} Our objective is to train a self-supervised fusion network that can reduce the influence of adversarial samples. The formal definition of this problem is as follows:
Here, we first define the dataset $Dataset$, which includes multiple user information, $Dataset = \{X_1,...,X_n\}$, and a face or voice verification model $V(X_o,\ X_i)$ in which $X_o$ is the original registered information and $X_i$ is the input user information. The output of the verification model $y = V(X_o,\ X_i)$ is a similarity score. Setting a threshold can give a prediction of whether they belong to the same person or not. Given an adversary $\mathcal{A}(V(X_i))$ that can access the verification model $V$, and it can produce an adversarial sample $X_a$ based on the input data. Our goal is to reduce the influence of the adversarial sample $X_a$ and make the prediction of $y = V(X_o,\ X_i)$ stable.

\subsubsection{Algorithm Details} The algorithm begins by initializing the necessary deep learning models, including the face and voice verification models, as well as the fusion defender ($\mathcal{D}$), which is designed to enhance security by combining multimodal inputs. The system continuously monitors incoming inputs, processing each pair of face and voice data to generate the respective embeddings. For each input pair, the voice recognition model ($VT$) extracts a subkey from the voice input, which is then used in combination with the face data to reconstruct both modalities via the fusion defender. This step ensures that both face and voice are considered in a unified verification process, reducing the risk of adversarial manipulation. Once the face and voice embeddings are extracted through their respective verification models, the system evaluates the authentication process by verifying if all three factors—face embedding, voice embedding, and the derived subkey—are consistent. If all three factors match, the delivery is accepted. Otherwise, the input is rejected, preventing unauthorized access.

\subsubsection{Model Architecture}

Here, we give the details of our proposed model. As shown in Figure \ref{fig:Fusion}, our proposed model also receives a sequence of flattened 2D patches as input like the other transformer-based models. We use two different linear projections for image and audio data to get the fixed-length embeddings. Position embeddings are added to the patch embeddings for the positional information. We use the same learnable 1D position embeddings in \cite{vit} and two image and audio data groups. After that, the embeddings are fed into the fusion transformer encoder, which consists of alternating layers of multiheaded self-attention and multilayer perceptron blocks. Then, we use a different linear projection for each embedding to reconstruct the original data. Specifically, we follow the base model of Vision transformer model variants. We use 12 layers of the transformer encoder and 12 heads for multi-head attention. The MLP dimension is 768, and we use dropout = 0.3.

\subsubsection{Loss Function}

Since our proposed model is a visual-audio fusion-based model, we combine two loss functions to train our network.

\textbf{Reconstruction Loss}. We use Mean Square Error $(MSE-loss)$ as the reconstruction loss:

\begin{equation}
\label{p}
MSE = \frac{1}{N}\sum_{i=1}^N{(y_i-\hat{y}_i)^2},
\end{equation}

where $N$ is the total number of data samples, and $y$ is the tensor value. 

\textbf{Overall Loss}. We have two types of loss functions for each type of data. The overall loss function that we train our network:

\begin{equation}
\label{p}
\mathcal{L}_{overall} =\lambda_{face}\mathcal{L}_{face} + \lambda_{voice}\mathcal{L}_{voice}
\end{equation}

where $\lambda_{face}$ and $\lambda_{voice}$ are regularization constants and $\mathcal{L}_{face}$ and $\mathcal{L}_{voice}$ are reconstruction loss of face and voice, respectively.

\subsection{Experimental Designs and Setups}
In our experiment, we aim to evaluate the resilience of our proposed defence model against adversarial samples in the context of face and voice verification models. 
All the experiments can be divided into two parts. The first part evaluates the robustness of the original face and voice verification model. The second part is employed to prove the effectiveness of our proposed fusion defender in comparison with several representative attack methods.


\subsubsection{Dataset}
We use the publicly available and widely used Voxceleb \cite{voxceleb} dataset for our experiments to train and test our proposed model. VoxCeleb contains over 100,000 voice and face data from 1251 different speakers. The dataset is divided into a training set of 1211 people and a test set of 40 people. We use the training and testing split from \cite{voxceleb} to train and test our model. 

\subsubsection{Attack Methodologies }
We leverage a wide range of attack methodologies (namely, FGSM \cite{goodfellow2014explaining}, PGD\cite{pgd}, BIM\cite{bim}, FFGSM\cite{ffgsm}, Jitter\cite{jitter}) with varying parameters
to ensure the generalizability of our proposed model. The perturbation levels are
selected such that the adversarial noise significantly influences the verification model.

\subsubsection{Evaluation Metrics}
To ensure a comprehensive evaluation of the defence model's performance, we utilized three metrics: accuracy, the Receiver Operating Characteristic (ROC) with the Area Under the Curve (AUC) value, and the Equal Error Rate (EER). Accuracy measures the defence model's overall prediction correctness, while the ROC curve, together with the AUC value, provides an integrated overview of the model's performance at various classification thresholds. The EER, on the other hand, offers an effective comparative metric for the balance between false acceptance and false rejection rates.
Our comprehensive evaluation methodology, involving multiple adversarial methods and diverse performance metrics, was designed to provide a detailed examination of the defence model's robustness against adversarial attacks, across both face and voice verification tasks. The results and discussion are provided in Section \ref{sec:discussion}.

\begin{figure*}[ht!]
\center{\includegraphics[scale=0.26]{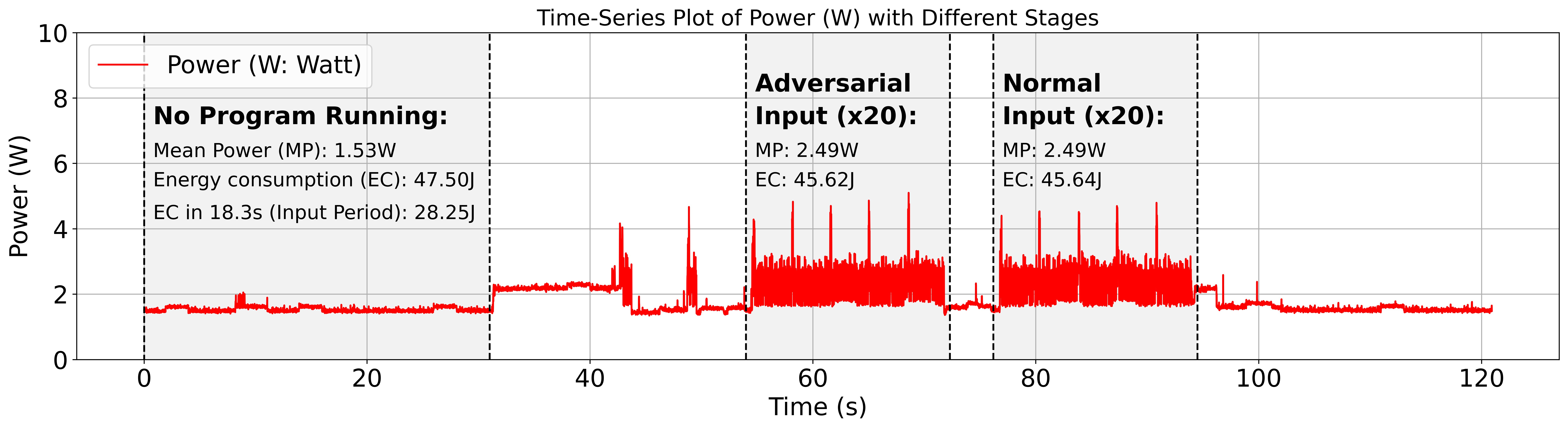} }
\caption{Power Consumption}
\label{fig:power} 
      \vspace{-4mm} 
\end{figure*}

\section{Discussion}
\label{sec:discussion}

In this section, we commence by presenting the specific details of our implementation of the proposed secure robotic delivery system. Subsequently, we demonstrate the computational expenses associated with achieving the crypto-factor of our multi-factor authentication phase. Following that, we proceed to evaluate the performance of the biometric-factor of our multi-factor authentication phase. Lastly, we validate the security of the crypto-factor component of our authentication phase through the utilization of two widely recognized symbolic-proof tools known as ProVerif and Scyther. We upload the codes and pre-trained models at Anonymous Google Drive\footnote{\textbf{Codes and pre-trained models are available at:} \url{https://drive.google.com/drive/folders/18B2YbxtV0Pyj5RSFX-ZzCGtFOyorBHil?usp=sharing}}. In this study, it is important to mention that we made the assumption that the robot successfully reached its destination by adhering to a predetermined path.

\vspace{-6mm} 

\subsection {Implemention Setup with the Computational Cost}

\begin{figure}
\center{\includegraphics[scale=0.1]{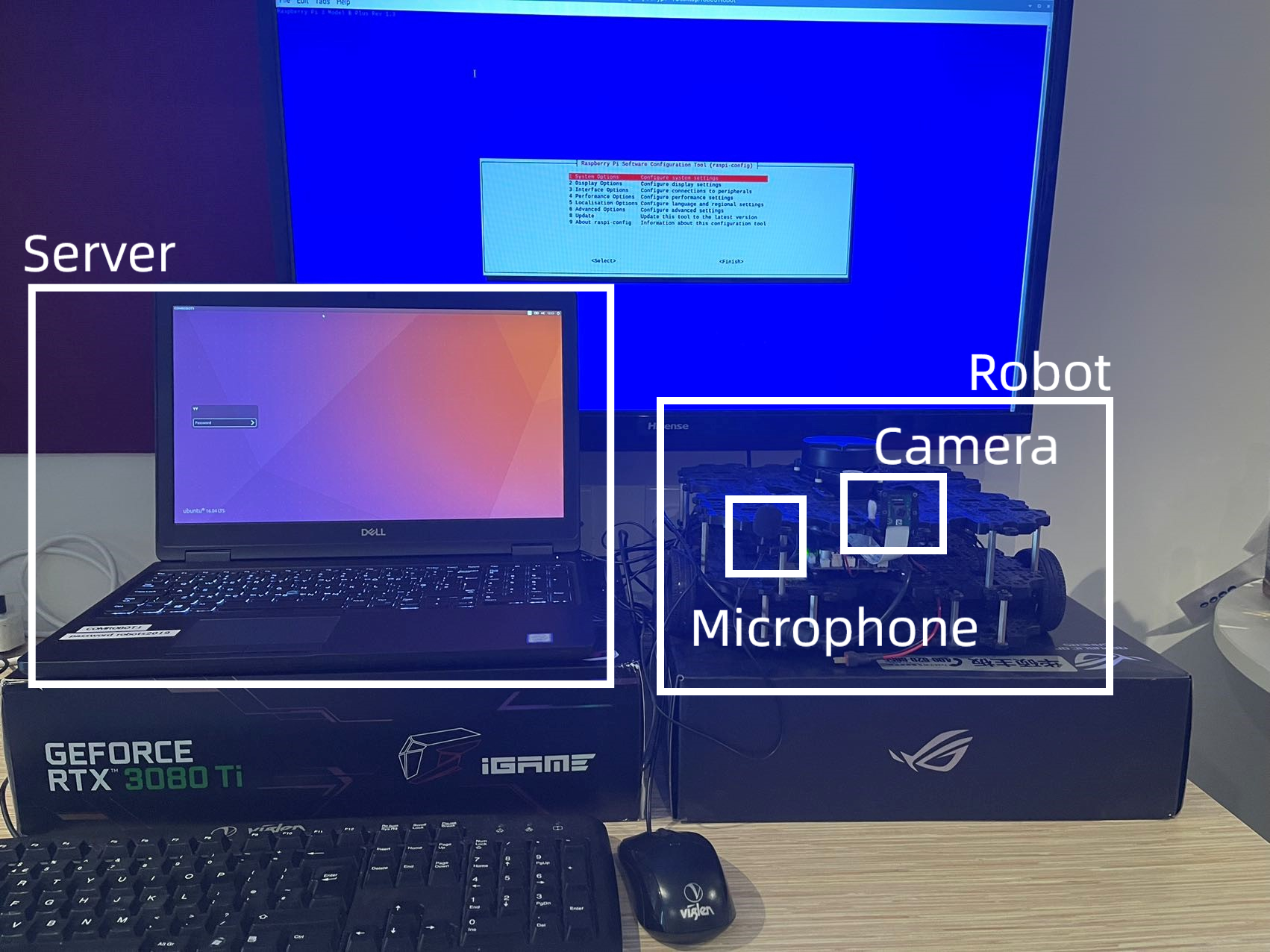} }
\caption{Implemention Setup}
\label{fig:Platform2} 
      \vspace{-6mm} 

\end{figure}

\begin{figure*}[htb]
\center{\includegraphics[scale=0.25,trim=0 0 0 0,clip]{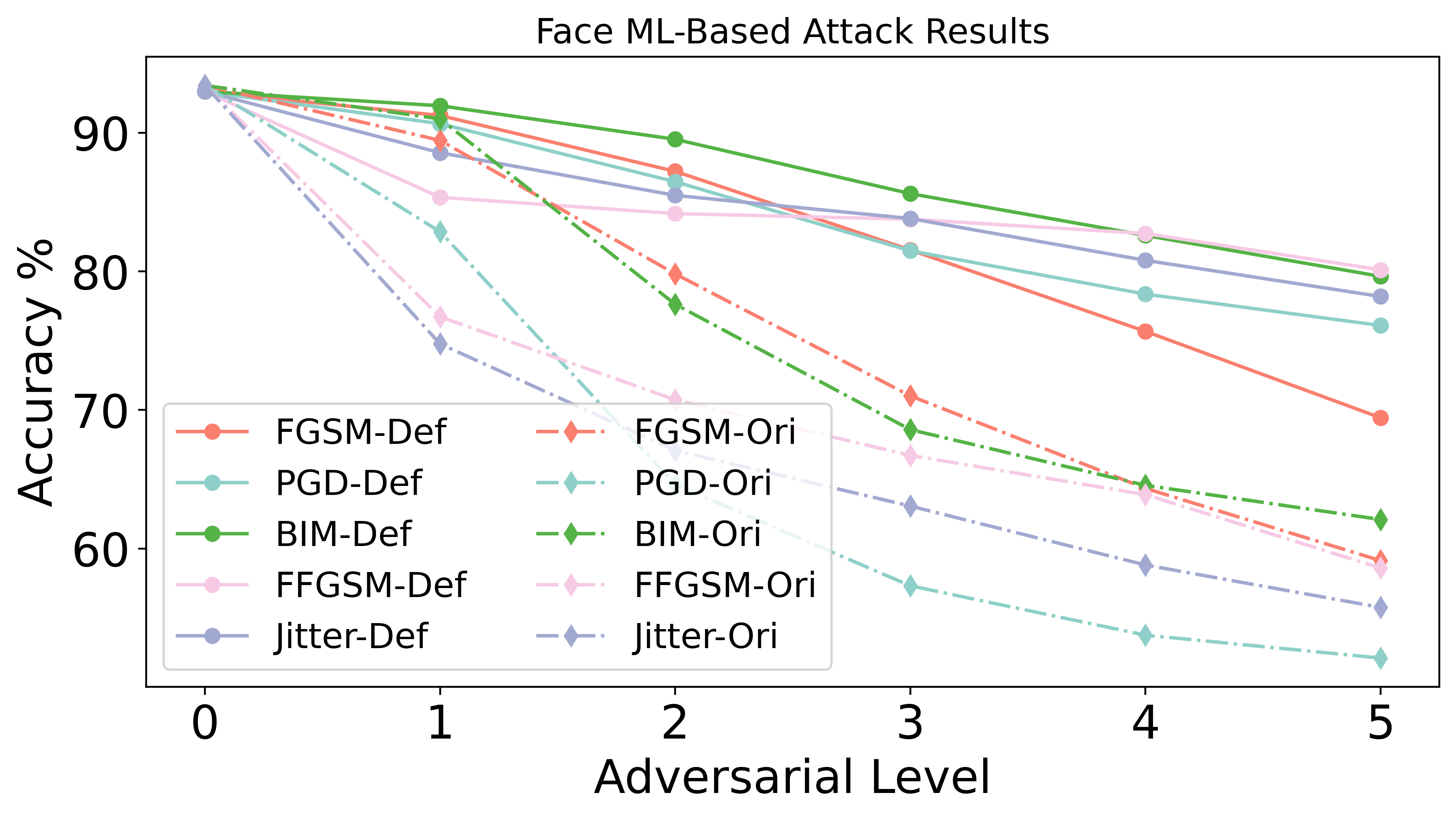}\includegraphics[scale=0.25,trim=0 0 0 0,clip]{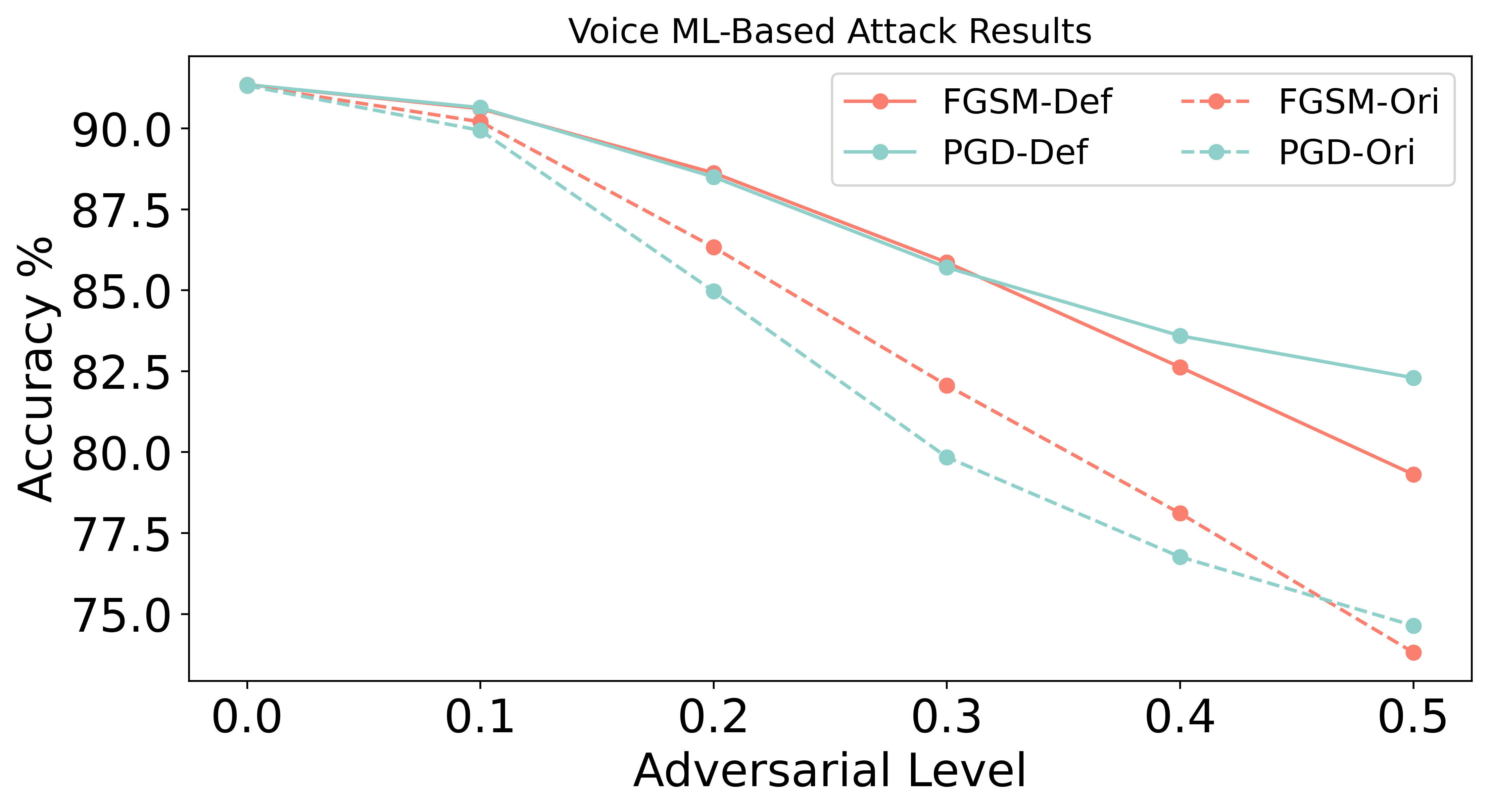}}
\caption{{Accuracy of ML-based Attack for Original and Defense Models}}
\label{fig:acc_ml} 
      \vspace{-6mm} 
\end{figure*}

In this section, we first describe the implementation details of our proposed secure robotic-delivery framework. Then, we will provide details on the computational cost for achieving the crypto-factor of our proposed authentication scheme.  Our experimental setup involved the utilization of two primary devices, namely the server and the robot. The server component was implemented on a Dell laptop equipped with a core i7, 2.30GHz CPU, and 16.0 GB RAM. The server operated on an Ubuntu 16.04 LTS system, utilizing Python 3.9.2 and PyTorch 1.12.0. For the execution of the robot delivery system, we employed the Turtlebot3 integrated with Raspberry Pi 3B+. The robot component was programmed using Python 3.5.2 and PyTorch 1.12.0. Figure \ref{fig:Platform2} illustrates the principal platforms employed in our experiment\footnote{\textbf{More videos and images can be accessed through the following links:} \url{https://drive.google.com/drive/folders/18B2YbxtV0Pyj5RSFX-ZzCGtFOyorBHil?usp=sharing}}.

We aimed to create a reliable and efficient experimental setup for evaluating our proposed protocol's performance and computational cost by employing these specific devices and software configurations. The experimental specifications are outlined in Table II in the Appendix, which includes information about the hardware configuration, computational cost, and communication specifications. Meanwhile, it presents computational specifications that offer meaningful insights into the execution times (in milliseconds and seconds) of various cryptographic operations utilized in the proposed scheme. The key generation and agreement process, facilitated by the ECDH operation, exhibits a computational cost of 0.458 s. The ECDSA operation, responsible for generating and verifying digital signatures, showcases a computational cost of 0.1.419 s for the setup phase and 0.053 s for the signing phase. Similarly, the encryption algorithm demonstrates a computational cost of 0.087s and the decryption algorithm, with a computational cost of 0.016s, for retrieving encrypted information. These computational costs provide valuable information into the performance and efficiency of the cryptographic operations, thereby contributing to a comprehensive evaluation of the proposed scheme's computational requirements.

Figure \ref{fig:power} shows the energy consumption by using a time series plot of Power (W: Watt). Due to the page limitation, we only provide the main results here. More detailed data can be found in the Anonymous Google Drive\footnote{\textbf{Data is available at:} \url{https://drive.google.com/drive/folders/18B2YbxtV0Pyj5RSFX-ZzCGtFOyorBHil?usp=sharing}}. We test the power consumption with three stages, which are: empty running (no program), adversarial input 20 times, and normal input 20 times. The mean power and energy consumption of empty running is 1.53W and 28.25J. The mean power of adversarial and normal input are both 2.49W, which means an adversarial attack won't cause extra workload for the model. Meanwhile, the energy consumption is close to each other, and for one input, it only costs 2.28J.

\begin{figure*}[htb]
\center{\includegraphics[scale=0.25,trim=0 0 0 0,clip]{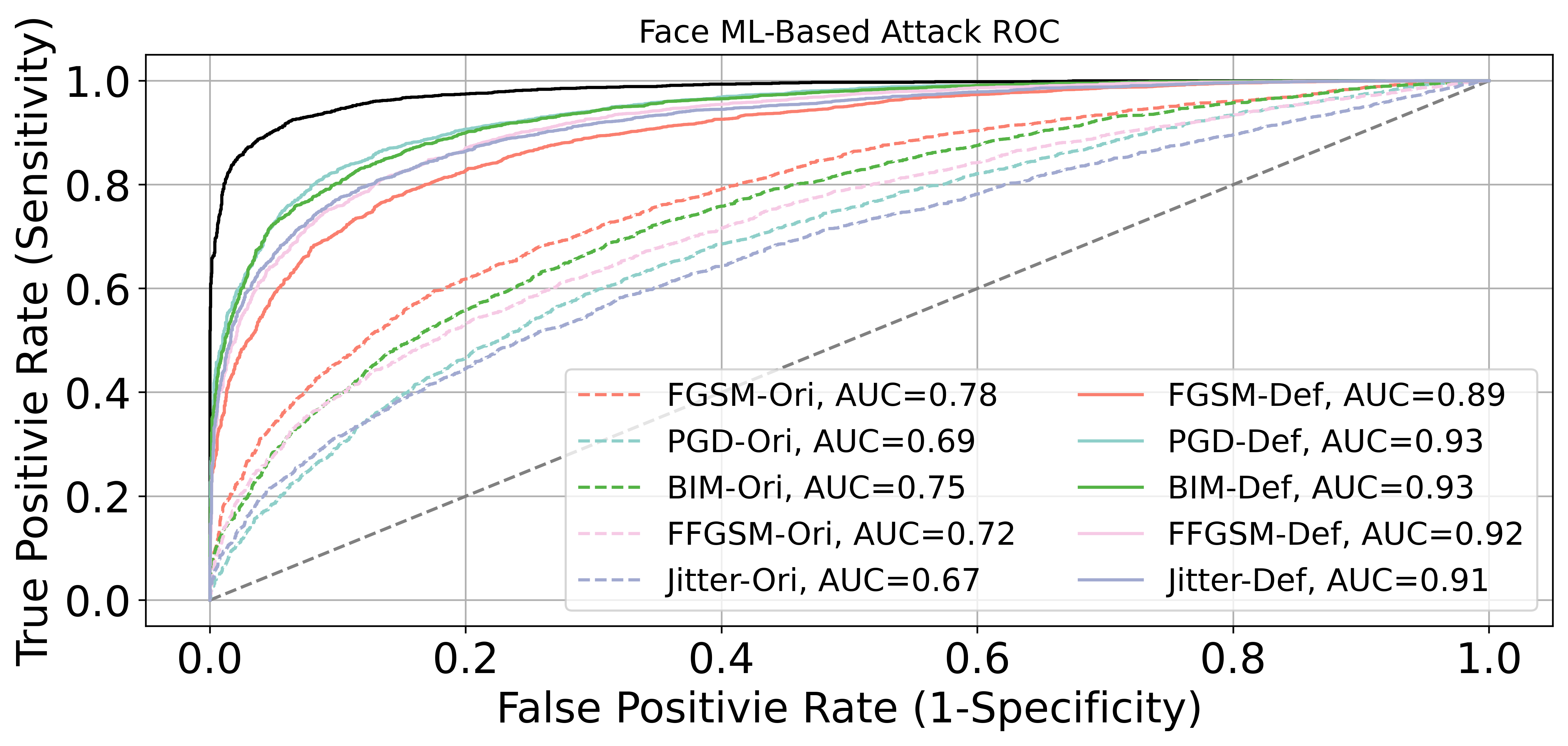}\includegraphics[scale=0.25,trim=0 0 0 0,clip]{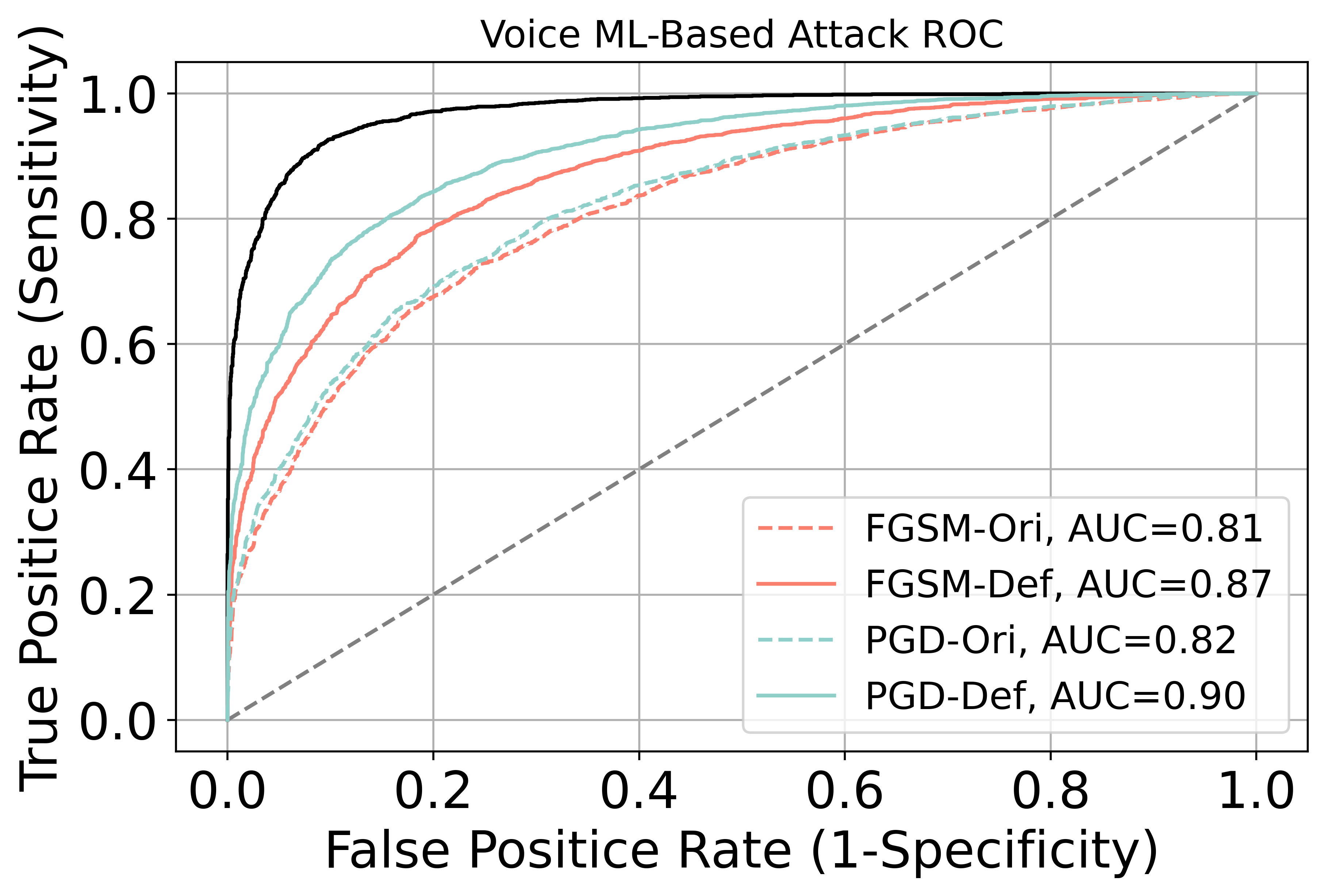}}
\caption{{ROC of ML-based Attack for Original and Defense Models}}
\label{fig:roc_ml} 
      \vspace{-5mm} 
\end{figure*}

\begin{table}[ht]
\caption{Equal Error Rate Result}
\centering
\begin{center}
\begin{tabular}{c c c c c c}
\toprule[1.5pt]
\multicolumn{6}{c}{{\textbf{Equal Error Rate Result Results (EER)}}} \\
\toprule[1.5pt]
\textbf{Attack Method} &  \makebox[0.05\textwidth] {\textbf{FGSM}} &  \makebox[0.05\textwidth] {\textbf{PGD}} &  \makebox[0.05\textwidth] {\textbf{BIM}} &  \makebox[0.05\textwidth] {\textbf{FFGSM}} &  \makebox[0.05\textwidth] {\textbf{Jitter}}  \\
\cmidrule[1pt](lr){1-6}
Original & 20.16 & 35.42 & 22.44 & 33.44 & 37.67 \\
Defence & \textbf{12.86} & \textbf{13.45} & \textbf{10.58} & \textbf{16.33} & \textbf{16.35} \\
\midrule[1.5pt]
\end{tabular}
\label{eer}
\end{center}
      \vspace{-10mm} 
\end{table}

\subsection{AI-based Biometric Security Analysis}

This section analyses the security of our proposed scheme's AI-driven Biometric Authentication Phase. In order to perform face verification and voice verification, we use ResNet to extract the user's face embeddings and ECAPA-TDNN \cite{desplanques2020ecapa} to extract the user's voice embeddings. Meanwhile, we use AAM-Softmax to get more accurate representations for users. To evaluate the resilience of our proposed defence model, we have utilized five different adversarial attack methods: FGSM, PGD, BIM, FFGSM, and Jitter. We present our evaluation outcomes using three metrics (discussed at \ref{sec:fusion}): Accuracy, ROC (Receiver Operating Characteristic), and EER (Equal Error Rate). We first start at Accuracy.

\subsubsection{\textbf{Accuracy of verification}}

Figure \ref{fig:acc_ml} shows the verification accuracy with different adversary levels. The results are divided into two categories: Ori and Def. Ori refers to the original model performance without any defence mechanism, while Def refers to the model performance with a defence mechanism in place. The horizontal axis represents the Adversarial Level, indicating the intensity or complexity of the adversarial attacks. The Accuracy \% on the vertical axis suggests the performance of the face verification system in terms of accuracy. The Def models for each attack type consistently perform better or at par compared to their Ori counterparts, indicating that the defence mechanism effectively mitigates the impact of the adversarial attacks.

\subsubsection{Receiver Operating Characteristic}
Figure \ref{fig:roc_ml} describes a ROC curve analysis of the performance of a face recognition system under various adversarial attacks. The ROC curves plot the True Positive Rate (TPR, or sensitivity) against the False Positive Rate (FPR, or 1-specificity) for different thresholds. The ROC curves for the original models (Ori) and the defended models (Def) are provided for each adversarial attack. The Ori models have dashed lines, while the Def models have solid lines. The area under the ROC curve (AUC) is calculated for each model, providing a summary measure of the model's performance.

An ideal ROC curve would reach the top left corner of the plot, implying high sensitivity (TPR) and low FPR at all thresholds. The black solid line represents an optimal model, while the grey dashed line represents the random guess line. Any model that lies above this line is considered better than random guessing. The Def models consistently have higher AUC values compared to their Ori counterparts for the same adversarial attack. This indicates that the defence mechanisms improve the robustness of the face recognition system against adversarial attacks.
 
\subsubsection{\textbf{Equal Error Rate}}
Table \ref{eer} provides an assessment of the EER performance of a defence model in contrast to the original model under various adversarial attack methods. For all the adversarial methods considered, it is evident that the defence model substantially outperforms the original model. This conclusion is drawn from a consistent decrease in the EER for the defence model across all attacks, indicating a more balanced trade-off between false acceptance and false rejection rates compared to their original counterparts. The largest improvement is witnessed under the PGD method, where the EER plunges from 35.42\% in the original model to 13.45\% in the defence model. Also, the results indicate a more balanced trade-off between false acceptance and false rejection rates compared to their original counterparts.

\subsubsection{\textbf{Resilience Against Physical Robot Capture Attack}} Although our proposed scheme has been designed to focus on digital security rather than the robot's physical security, however in the event of any physical attack on the robot, our proposed scheme can still protect sensitive user data, such as facial and voice recognition information. In this regard, we do not directly store any biometric information (such as the user's facial image or voice) at the robot's side. The information stored on the robot side is extracted from the deep learning model (face embedding and voice embedding). Meanwhile, the information is also encrypted using a one-time key generated by the server. If we assume that the attacker can break the one-time key from the server side, since our proposed protocol ensures forward and backward secrecy (as shown in the Appendix). In that case, our scheme can still guarantee the security of any previous sessions. In this way, in the event of a physical attack by a robot, we can still guarantee the security and privacy of the whole system.

\vspace{-3mm}

\section{Conclusion}
\label{sec:Conclusion}

In this paper, we propose a new privacy-preserving multi-factor authentication scheme for secure robotic-based delivery systems. Unlike prior works, we combine a multi-factor authentication scheme with a secure cryptographic protocol to ensure the robotic delivery system's desirable security and privacy properties. This protocol can also be applied to other scenarios, like medicine delivery with sensitive information.
 While providing the security protocol, we also proposed the first audio-visual fusion transformer-based defender against the adversarial samples aimed at AI-assisted systems. Since deep learning-based authentication is an essential component in modern biometric verification, it is also significant to ensure the security of such models.

\vspace{-3mm}
 
\bibliographystyle{IEEEtran} 
\bibliography{main}

\end{document}